\documentclass[twocolumn,pof,superscriptaddress,showpacs]{revtex4}
\usepackage{epsf,graphicx,amssymb,subfig,amsmath,color}
\begin{document}
\title{The role of boundaries in the MagnetoRotational Instability}

\author{Christophe Gissinger} \affiliation{Department of Astrophysical
  Sciences, Princeton University, Princeton, NJ
  08544.}\affiliation{Center for Magnetic Self-Organization in
  Laboratory and Astrophysical Plasmas, Princeton Plasma Physics
  Laboratory, Princeton University, P.O. Box 451, Princeton, New
  Jersey 08543, USA} \author{Jeremy Goodman} \affiliation{Center for
  Magnetic Self-Organization in Laboratory and Astrophysical Plasmas,
  Princeton Plasma Physics Laboratory, Princeton University, P.O. Box
  451, Princeton, New Jersey 08543, USA} \author{Hantao Ji}
\affiliation{Department of Astrophysical Sciences, Princeton
  University, Princeton, NJ 08544.}

\newcommand{\bfnabla}{\boldsymbol{\nabla}} 
\newcommand{\bB}{\boldsymbol{B}}                   
\newcommand{\bj}{\boldsymbol{j}}                     
\newcommand{\bu}{\boldsymbol{u}}                  
\newcommand{\btimes}{\boldsymbol{\times}}   
\newcommand{\remark}[1]{{\color{red}\bf #1}} 

\begin{abstract}
In this paper, we investigate numerically the flow of an electrically
conducting fluid in a cylindrical Taylor-Couette flow when an axial
magnetic field is applied. To minimize Ekman recirculation due to
vertical no-slip boundaries, two independently rotating rings are used
at the vertical endcaps. This configuration reproduces setup used in
laboratory experiments aiming to observe the MagnetoRotational
Instability (MRI). Our 3D global simulations show that the nature of
the bifurcation, the non-linear saturation, and the structure of
axisymmetric MRI modes are significantly affected by the presence of
boundaries. In addition, large scale non-axisymmetric modes are
obtained when the applied field is sufficiently strong. We show that
these modes are related to Kelvin-Helmoltz destabilization of a free
Shercliff shear layer created by the combined action of the applied
field and the rotating rings at the endcaps. Finally, we compare our
numerical simulations to recent experimental results obtained in the
Princeton MRI experiment.

\end{abstract}
\maketitle 

\section{Introduction}

The magnetorotational instability has provided a simple explanation of
the longstanding problem of rapid angular momentum transport in
accretion disks around stars and black
holes~\cite{BalbusHawley91}. Balbus and Hawley, rediscovering an
instability first studied by Velikhov \cite{Velikhov59} and
Chandrasekhar \cite{Chandra60}, have shown that keplerian flows of
accretion discs, for which the Rayleigh criterion predicts axisymmetric
hydrodynamical stability, can be destabilized in the presence of a
magnetic field. More precisely, the MRI is a linear instability
occurring when a weak magnetic field is applied to a rotating
electrically conducting fluid for which the angular velocity decreases
with the distance from the rotation axis,
i.e. $\partial_R(\Omega^2)<0$. Linear stability analysis indicates
that the most unstable mode is axisymmetric and associated with a
strong radial outflow of angular momentum. Non-linear evolution of the
MRI is of primary importance, since saturation of the instability
eventually yields a magnetohydrodynamical turbulent state, enhancing
the angular momentum transport.\\

During the last decade, there has been a lot of effort to observe the
MRI in a laboratory experiment. To this end, most groups use a
magnetized Taylor-Couette flow, i.e. the viscous flow of an
electrically conducting fluid confined between two differentially
rotating concentric cylinders, in the presence of an externally imposed
magnetic field. For infinitely long cylinders, the ideal laminar
Couette solution is given by:
\begin{equation}
\Omega(r)=A_1+\frac{A_2}{r^2}
\label{theoCouette}
\end{equation}
in which $A_1=(\Omega_2r_2^2-\Omega_1r_1^2)/(r_2^2-r_1^2)$ and
$A_2=r_1^2r_2^2(\Omega_1-\Omega_2)/(r_2^2-r_1^2)$, $\Omega_1$ and
$\Omega_2$ are respectively the angular velocity of inner and outer
cylinder, and $r_1$, $r_2$ are the corresponding radii.
The Rayleigh criterion predicts axisymmetric linear stability if
$\Omega_2/\Omega_1 \ge (r_1/r_2)^2$, ensuring that the 
specific angular momentum increases outward.
However, MRI may arise if $\Omega_2/\Omega_1<1$ and non-ideal
MHD effects are small \cite{Ji01}.

Several experiments are currently working on MRI. The Princeton
experiment has been designed to observe this instability in a
Taylor-Couette flow of liquid Gallium, with an axial applied magnetic
field \cite{Ji06}. So far, MRI has not been identified, but
non-axisymmetric modes have been observed when a strong magnetic field
is imposed \cite{Nornberg10}. The PROMISE experiment, in Dresden, is
based on a similar set-up, except that the applied field possesses an
azimuthal component. Axisymmetric traveling waves have been obtained,
and identified as being Helical MRI, an inductionless instability
different from but connected to the standard MRI \cite{Stefani06}.  A
few years ago, MRI was claimed to have been obtained in a
spherical Couette flow of liquid sodium (hereafter ``Maryland experiment'') \cite{Sisan04}. In
this experiment, in which the outer sphere was at rest and a
poloidal magnetic field was applied, non-axisymmetric oscillations of both
velocity and magnetic fields were observed, together with an
increase of the torque on the inner sphere. However, it has been
shown recently that these oscillations were more likely related to
instabilities of magnetic free shear layers in the flow
\cite{Gissinger11}, \cite{Hollerbach09}.\\

Experimental observation of the MRI is
considerably complicated by the presence of vertical
boundaries. Indeed, no-slip boundary conditions at the vertical
endcaps (for instance rigidly rotating with either the outer or the
inner cylinder) induce an imbalance between the pressure gradient and
the centrifugal force, and drive a meridional Ekman recirculation in addition to
the azimuthal Couette flow.
An inward boundary-layer flow near the endcaps is balanced by
strong outward jet at the midplane. This Ekman recirculation has two
important consequences for laboratory MRI: first,
the meridional flow transports angular momentum outward, decreasing
the free energy available to excite MRI. Moreover, the
Ekman flow is a source of hydrodynamical fluctuations,  
beginning at Reynolds numbers $\gtrsim 400$ with oscillations of the radial jet \cite{Kageyama04},
which strongly complicates the identification of MRI modes.

In the Princeton MRI experiment, this problem has been circumvented by
replacing the rigid endcaps at the top and the bottom by two rings
that are driven independently, reducing the imbalance due to viscous
stress. It has been shown that a quasi-keplerian flow profile 
$(r_1/r_2)^2<\Omega_2/\Omega_1<1$ 
in a short
Taylor-Couette cell can be kept effectively stable up to $Re\sim 10^{6}$ if
appropriate ring speeds are used
\cite{Ji06}. Similarly, the PROMISE group found that
 split rings (attached to the cylinders rather
than independently driven) led to a significant reduction of the Ekman pumping and
a much clearer identification of the HMRI \cite{Stefani09}.\\

In this article, we report three-dimensional numerical simulations
inspired by these experimental configurations, especially the
Princeton MRI experiment. In the first section, we present the
equations and the numerical method used. In section II, we study how
the structure and the saturation of the magnetorotational instability
arising in Taylor-Couette flow is modified by the presence of vertical
boundaries. In section III, we report new non-axisymmetric
instabilities generated by a free shear layer in the flow when the
applied field is sufficiently strong compared to the
rotation. Finally, a comparison with results from the Princeton MRI
experiment is presented.

\section{ Equations}
We consider the flow of an electrically conducting fluid between two
co-rotating finite cylinders. $r_1$ is the radius of the inner
cylinder, $r_2=3r_1$ is the radius of the outer cylinder, and
$d=r_2-r_1$ is the gap between cylinders. $\Omega_1$ and $\Omega_2$
are respectively the angular speed of the inner and the outer
cylinder. The height of the cylinders $H$ is fixed such that we have
the aspect ratio $\Gamma=H/d=2$, as in the Princeton MRI
experiment.  A uniform background magnetic field $\bB_0=B_0\boldsymbol{e}_z$ is
imposed by external coils.

The governing equations for this problem are the Navier-Stokes
equations coupled to the induction equation :
 \begin{equation}
 \rho\frac{\partial \bu}{\partial t} + \rho\left( \bu\cdot\bfnabla\right)  
\bu=-\bfnabla P+\rho\nu\bfnabla^{2}\bu +\bj\btimes\bB  \ .
\label{NS}
\end{equation}
\begin{equation}
\frac{\partial\bB}{\partial t}=\bfnabla\btimes\left(\bu\btimes\bB\right)+\frac{1}{\mu_{0}\sigma}\bfnabla^2\bB  \ .
\label{ind}
 \end{equation}
where $\rho$ is the density, $\nu$ the kinematic viscosity,
$\eta=1/(\sigma\mu_0)$ is the magnetic diffusivity, $\bu$ is the fluid
velocity, $\bB$ the magnetic field, and $\bj=\mu_0^{-1}
\bfnabla\btimes\bB$ the electrical current density.  The problem is
characterized by three dimensionless numbers: the Reynolds number
$Re=(\Omega_1 r_1d)/\nu$, the magnetic Reynolds number $Rm=(\Omega_1
r_1d)/\eta$ and the Lundquist number
$S=B_0r_1/(\sqrt{\rho\mu_0}\eta)$.  Alternatively, the applied field
can be measured by the Hartmann number
$M=B_0r_1/\sqrt{\rho\mu_0\eta\nu}$, or the Elsasser number
$\Lambda=B_0^2/(\rho\mu_0\eta\Omega)$. It is also useful to define the
magnetic Prandtl number $Pm=\nu/\eta$, which is simply the ratio
between Reynolds numbers.  At the top and bottom, the endcaps are
divided at $r=(r_1+r_2)/2$ between two independently concentric
rotating rings. Rotation rates of inner and outer rings are given
respectively by $\Omega_3$ and $\Omega_4$.\\

These equations are integrated with the HERACLES code
\cite{Gonzales06}. Originally
developed for radiative astrophysical compressible and ideal-MHD
flows, HERACLES relies on a finite volume Godunov method. The code has been
parallelized with the MPI library and implemented in Cartesian,
cylindrical and spherical coordinates. We have modified this code to
include fluid viscosity, magnetic resistivity, and suitable boundary
conditions for velocity and magnetic fields. Note that HERACLES
is a compressible code, whereas laboratory experiments generally
involve almost incompressible liquid metals. In fact,
incompressibility corresponds to an idealization in the limit of
infinitely small Mach number ($Ma$).  In the
simulations reported here, we used an isothermal equation of state
with a sound speed such that $Ma<0.1$,
following the approach of \cite{Liu06},\cite{Liu08}. Typical
resolutions used in the simulations reported in this article are
$(N_r,N_\phi,N_Z)=[200,64,400]$. For some runs, we have checked that
doubling this resolution does not affect results.\\

Except where indicated, the Reynolds number has been
fixed at $Re=4000$, whereas $Rm$ and $S$ vary widely.
On the cylinder sidewalls, at $r=r_1$ and $r=r_2$, we impose
perfectly electrically conducting boundaries :

\begin{equation}
B_r=0 \ , \hspace{15pt}
{\partial B_z   }/{\partial r}=0\ , \hspace{15pt}
{\partial B_\phi   }/{\partial r}=0
\end{equation}

At the vertical endcaps, we use the so-called pseudo-vacuum boundary
conditions, namely the magnetic field is forced to be normal to the
boundary:
\begin{equation}
B_r=0 \ , \hspace{15pt}
B_\phi=0 \ , \hspace{15pt}
{\partial B_z}/{\partial z}=0 
\label{fero}
\end{equation}
With these conditions, no \emph{magnetic} torques are exerted through
the boundaries.  For the velocity field, no-slip boundary conditions
are used for radial and axial components. The angular velocity matches
the rotation rates of cylinders and rotating rings at radial and axial
boundaries.

\section{ Magneto-rotational instability in finite geometry}

\subsection{Purely hydrodynamical simulations}

Accretion discs result from the balance between gravitation and
centrifugal force, leading to an angular velocity profile such that
$\Omega\sim r^{-3/2}$. Since the Rayleigh criterion states that any
inviscid flow with rotation profile $\Omega\sim r^{\delta}$ is
linearly stable to the centrifugal instability if $\delta>-2$,
accretion discs are expected to be hydrodynamically linearly
stable. Although Keplerian profiles can not be reproduced in the
laboratory, it is possible to drive a Taylor-Couette flow in the
so-called {\it quasi-keplerian} regime, i.e. with an angular momentum
which increases outward to satisfy Rayleigh stability criterion, but
which is MRI unstable (the angular velocity decreases outward). Except
where indicated, all simulations reported here use the following
rotation rates for the cylinders and rings:
\begin{equation}
\Omega_1=1. \ , \hspace{10pt}
\Omega_2=0.13 \ , \hspace{10pt}
\Omega_3=0.45 \ , \hspace{10pt}
\Omega_4=0.16 \ 
\label{Ome1}
\end{equation}
where rotation rates have been normalized by $\Omega_1$. The ratio
$\Omega_2/\Omega_1$ ensures a quasi-keplerian regime, whereas
$\Omega_3$ and $\Omega_4$ have been carefully chosen to strongly
reduce Ekman recirculation and therefore reinforce the hydrodynamical
stability of the flow \cite{Liu08}.
\begin{figure}
\centerline{
\epsfysize=50mm 
\epsffile{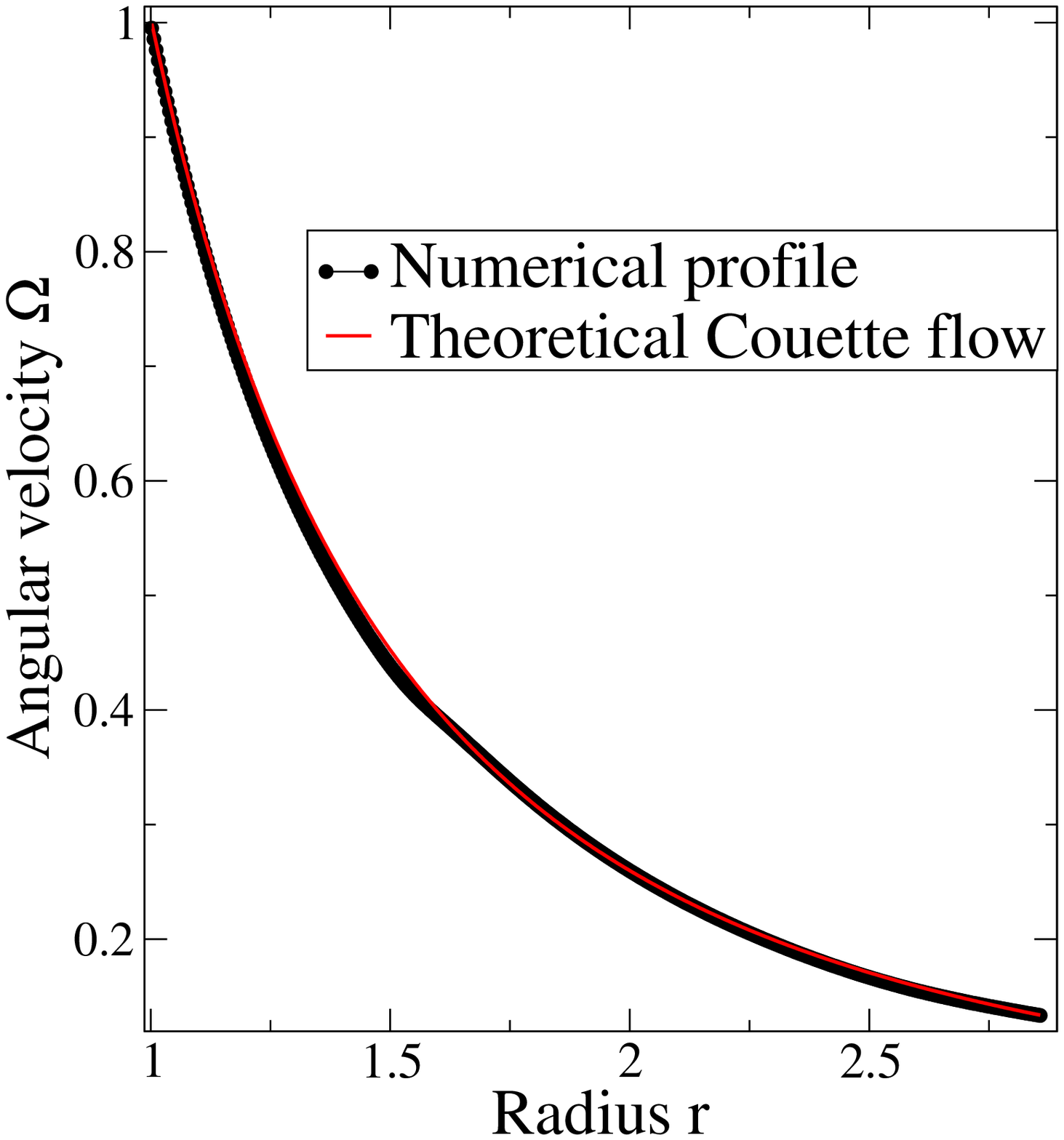} 
\epsfysize=53mm 
\epsffile{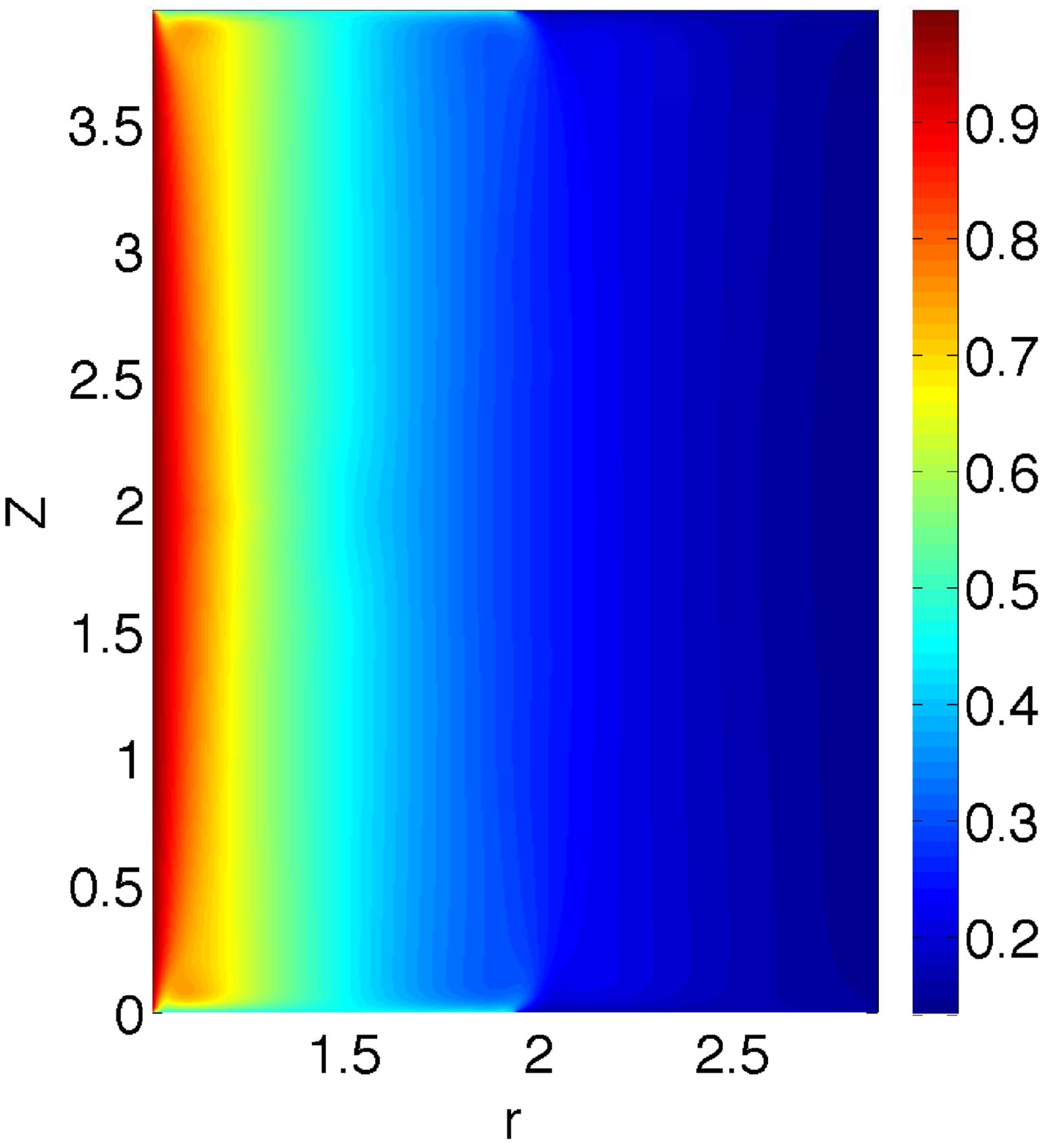} }
\centerline{
\epsfysize=53mm 
\epsffile{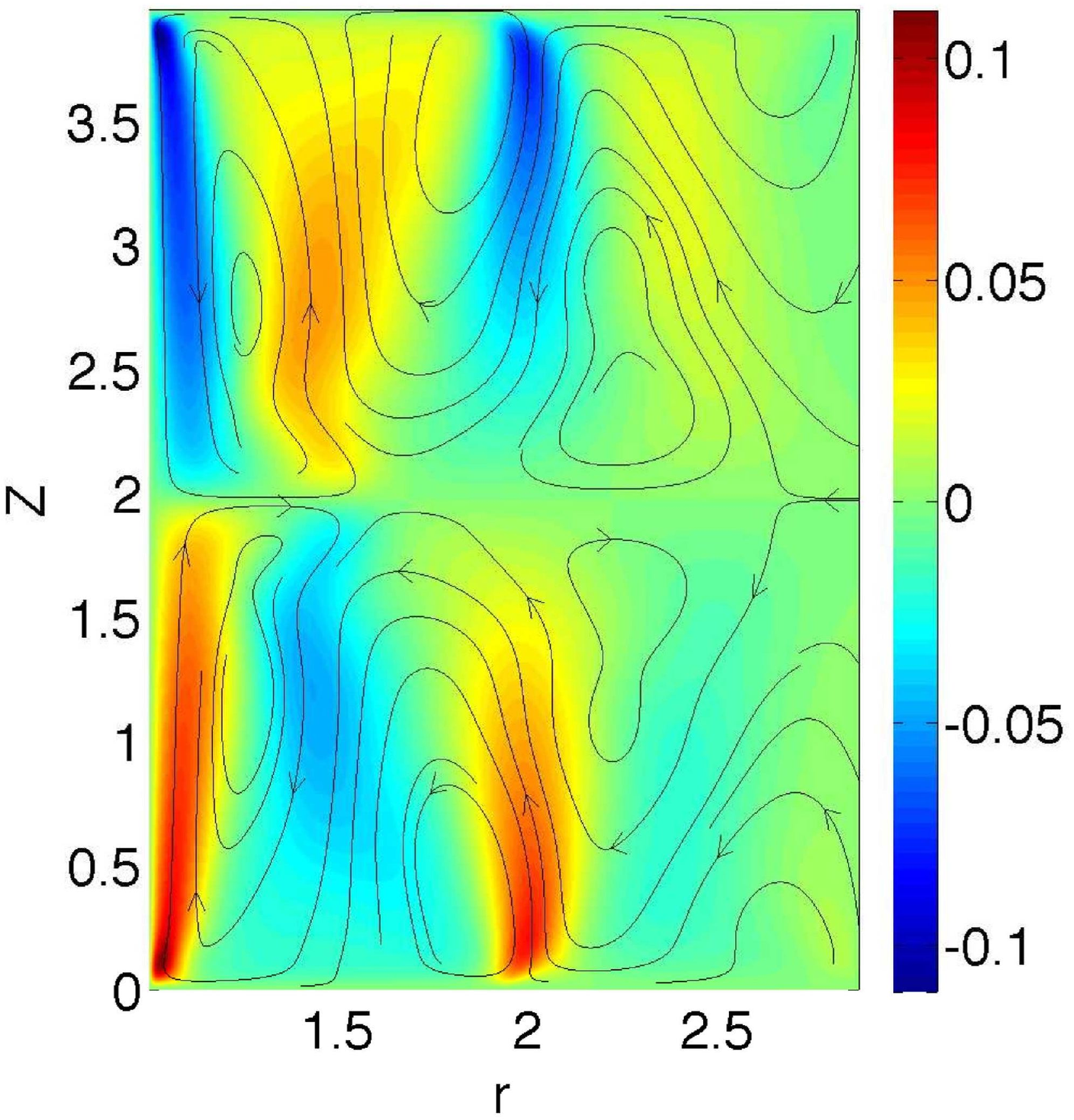} 
\epsfysize=53mm 
\epsffile{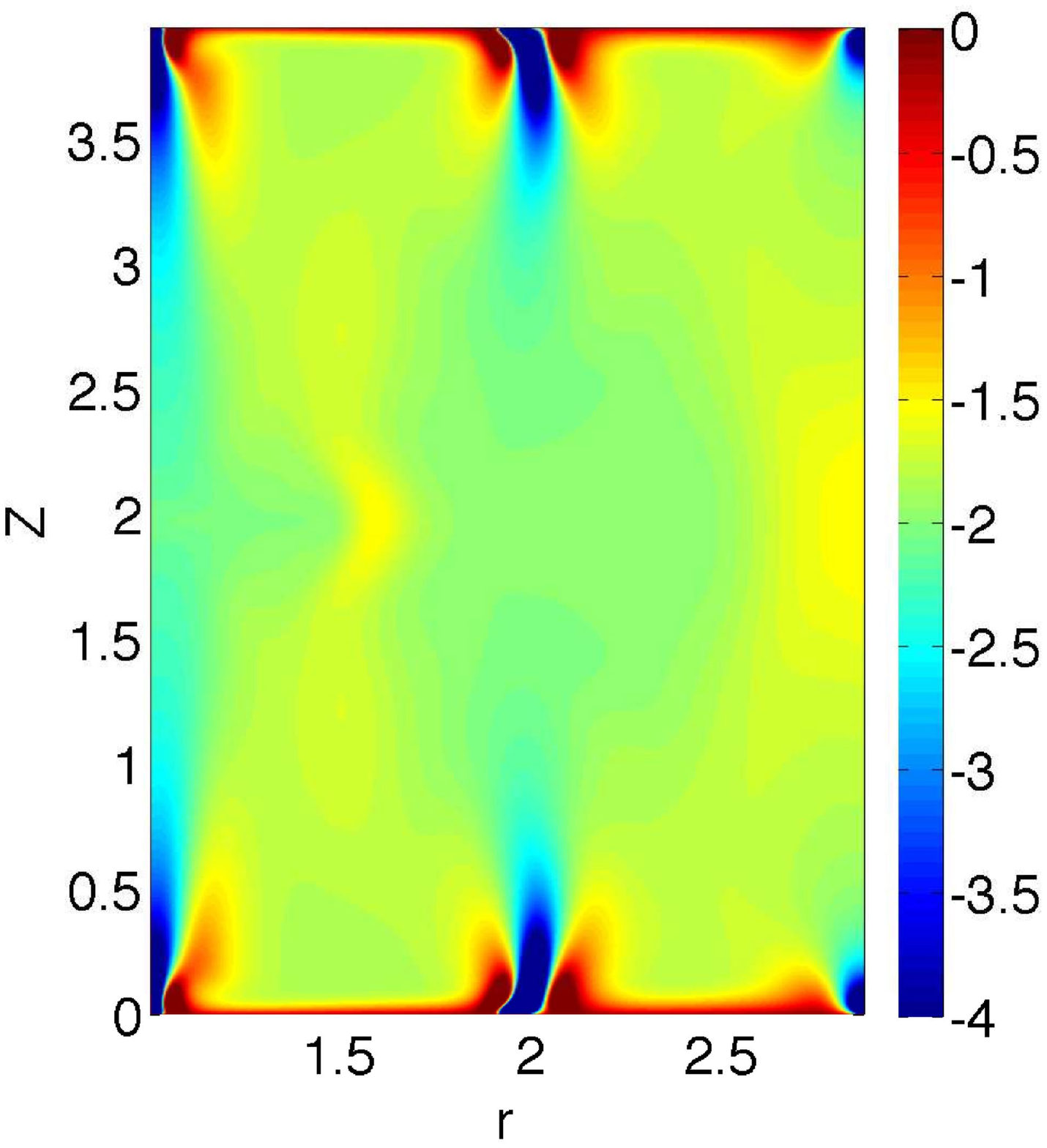} }
\caption{ Purely hydrodynamical axisymmetric Taylor-Couette flow,
  obtained for $Re=4000$, $Rm=15$, $S=0$ and rotations rates given by
  eq.(\ref{Ome1}). Top, left: profile of $\Omega$ at $z=h/4$. Top,
  right: angular velocity $\Omega$ in $(r,z)$ plane. Bottom, left:
  $U_z$ in the $(r,z)$ plane (streamlines indicate meridional
  flow). Bottom, right: Shear $q$ (see eq. (\ref{shear_eq}) ) in the
  $(r,z)$ plane. The use of rings at endcaps allows the flow to be
  very close to theoretical circular Couette flow. There is a weak
  poloidal recirculation, with $8$ unsteady cells. Except close to the
  boundaries, the shear is relatively uniform.}
\label{Couette}
\end{figure}

Figure \ref{Couette} shows the flow obtained with the aforementioned
standard parameters but without magnetic field, and for an
axisymmetric simulation. Note that the radial profile of angular
velocity matches almost perfectly the theoretical Couette solution
(Fig.~\ref{Couette}, top-left) that would be obtained with infinite or
$z$-periodic cylinders. In fact, figure \ref{Couette},top-right shows
that the Couette profile prevails not only in the midplane, but in
most of the domain. As expected, the use of rotating rings at the
endcaps strongly reduces the Ekman flow, dividing the usual two big
Ekman cells into eight smaller and less intense cells
(Fig.~\ref{Couette},bottom-left). This structure is unsteady and the
corresponding weak poloidal flow fluctuates as time evolves. The
amplitudes of $u_r$ and $u_z$ are one order of magnitude smaller than
for endcaps rotating with the outer cylinder.

To compare our numerical results to theoretical predictions, it is
useful to define a local shear parameter $q$:
\begin{equation}
q=\frac{r}{\Omega(r)}\frac{\partial \Omega(r)}{ \partial r}
\label{shear_eq}
\end{equation}
For instance, in a keplerian flow $q$ would have the constant value
$-3/2$.  Figure~\ref{Couette},bottom-right shows $q$ in the $(r,z)$-plane for our
system. In the bulk of the flow, $q$ is between $-1.5$ and $-2$ and
matches the value corresponding to an ideal circular Couette
flow. However, close to the endcaps, a very strong shear is generated
where the rings meets, characterized by $q<-3.5$, reflecting
Rayleigh-unstable flow. This corresponds to a cylindrical Stewartson
free shear layer \cite{Stewartson66} at $r=(r_1+r_2)/2$, created by
the jump of the angular velocity between the two independently
rotating rings. See \cite{Liu08b} for a more detailed numerical study
of this layer. Although one could expect a destabilization of this
shear layer to large scale non-axisymmetric modes \cite{Hollerbach05},
this has not been observed in our simulations. Note that the layer
does not extend very far from the endcaps, suggesting that it is
disrupted by small-scale instabilities.\\

\begin{figure}
\centerline{
\epsfysize=80mm 
\epsffile{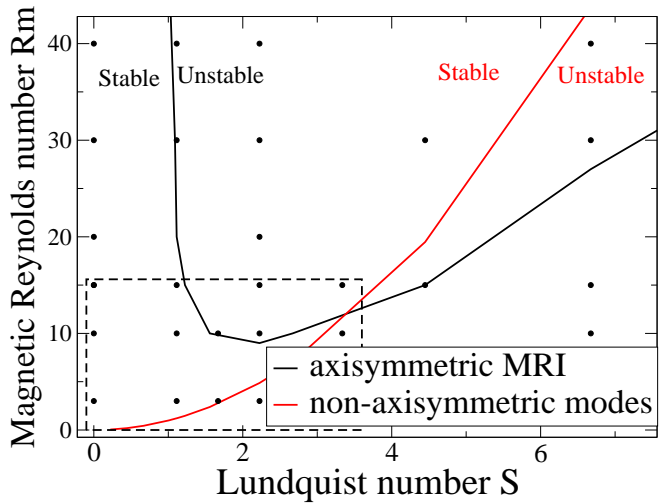} }
\caption{Stability boundaries in the plane of magnetic Reynolds number, $Rm$,
versus  Lundquist number, $S$. Black curve
  indicates the marginal stability curve of the MagnetoRotational
  Instability (MRI) obtained by interpolation from $2D$
  simulations. Red curve is the marginal stability curve of
  non-axisymmetric modes generated by Kelvin-Helmoltz destabilization
  of the MHD detached shear layer, obtained from $3D$ simulations
  (black dots). Dashed square indicates the parameter space
  accessible to the Princeton experiment. }
\label{curve}
\end{figure}

\subsection{Imperfect bifurcation of the magnetorotational instability}

Let us now consider the magnetized problem, meaning that a homogeneous
axial magnetic field is now applied at the beginning of the
simulation. As expected from global and local linear analyses, the
flow is destabilized to MRI for sufficiently large $Rm$ and
$S$. Figure \ref{curve} shows the marginal stability curve of the MRI
(black curve) in the plane defined by these parameters. In the domain
explored here, MRI modes are always axisymmetric. Linear stability
analysis predicts non-axisymmetric MRI modes for $Rm>50$, which is
larger than considered here. Note that the non-axisymmetric modes in
Fig. \ref{curve} are not standard MRI but are related to the
instability described in the next section. If the applied field is too
strong, MRI modes are suppressed, and the flow is restabilized to a
laminar state dominated by the azimuthal velocity. As a consequence,
MRI modes are unstable only within the pocket-shaped interior of the
black curve.\\

\begin{figure}
\centerline{
\epsfysize=50mm 
\epsffile{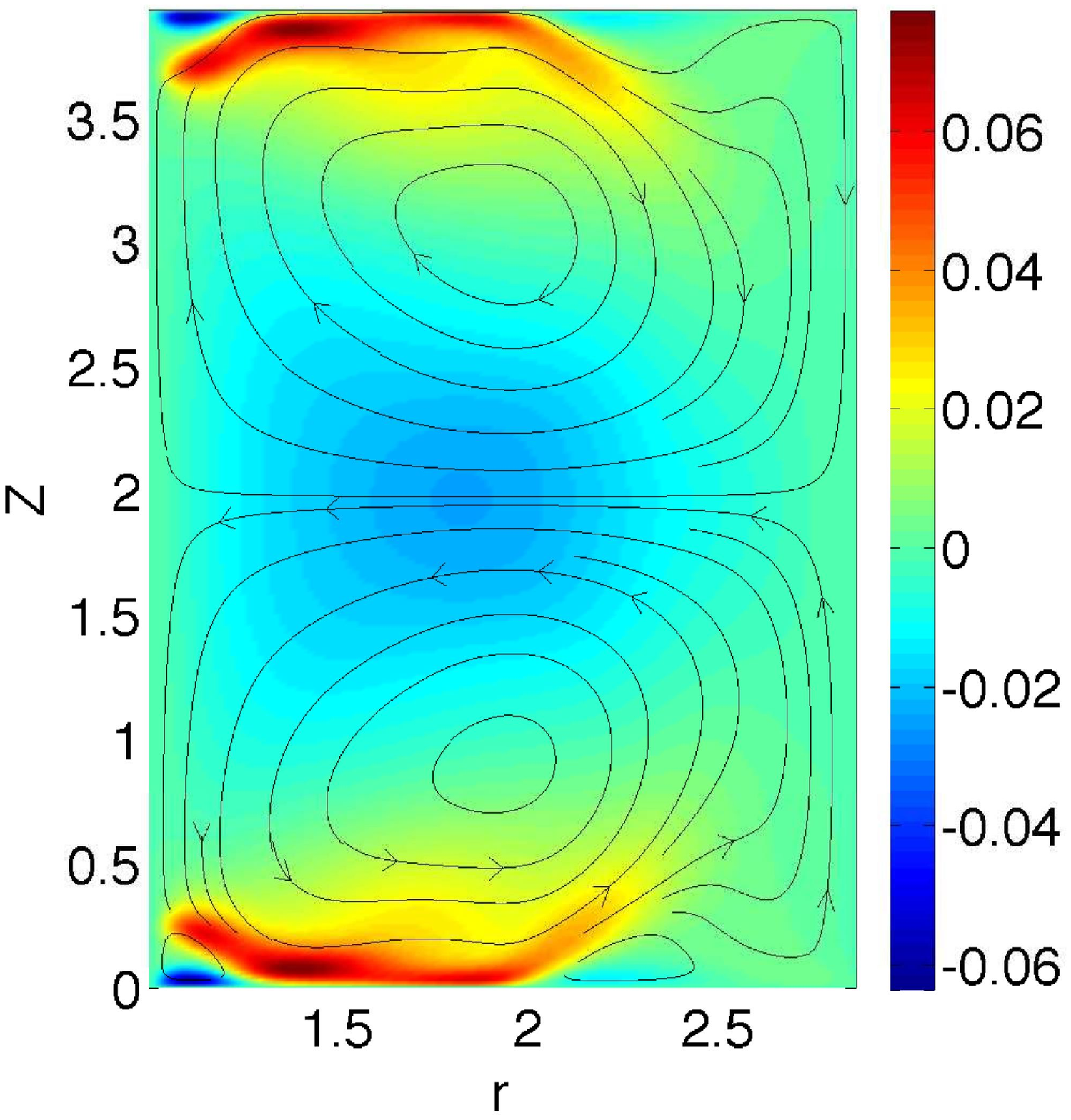} 
\epsfysize=50mm 
\epsffile{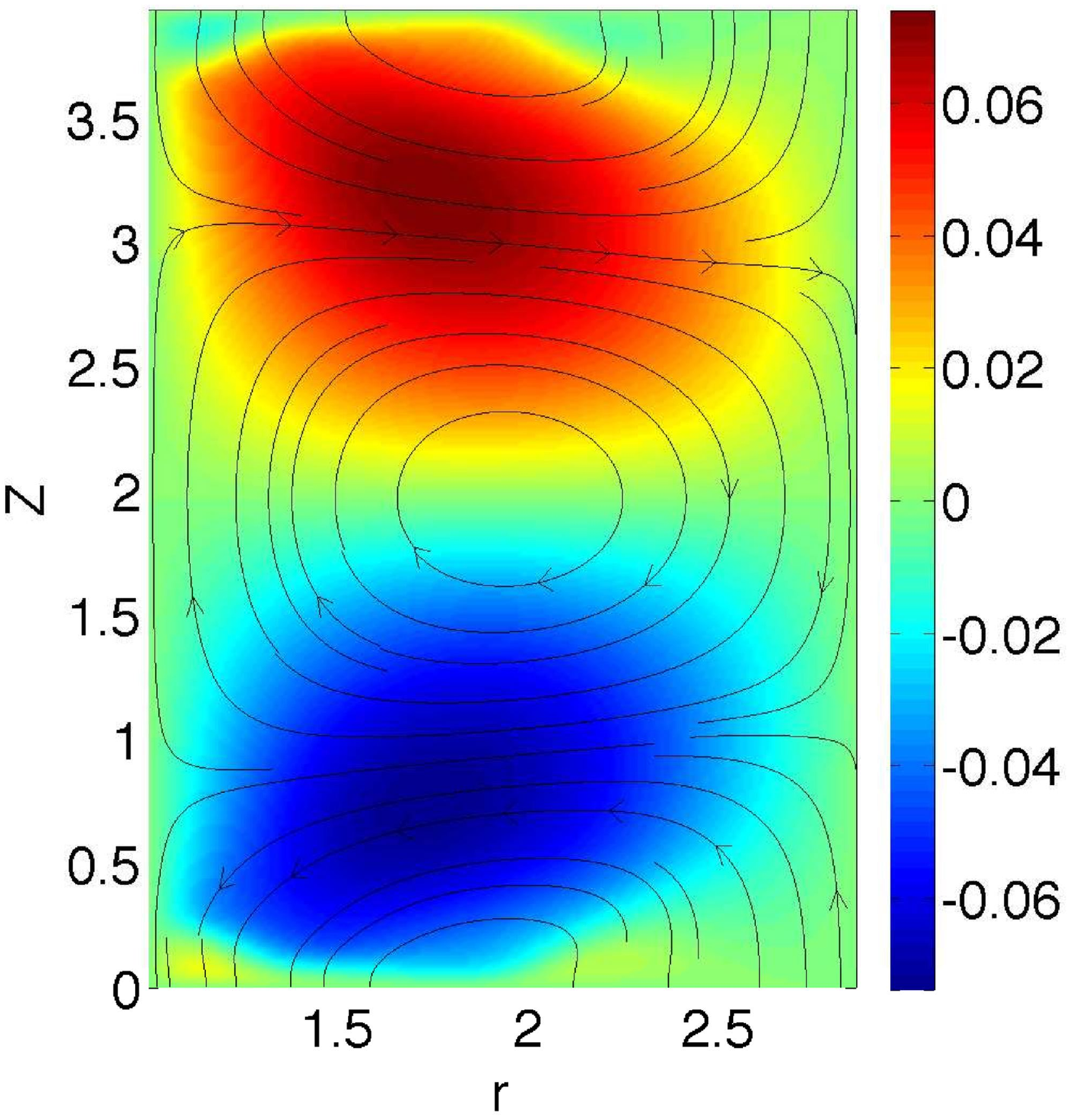} }
\caption{ Structure of the MagnetoRotational Instability (MRI) mode
  when a magnetic field is applied in the $z$-direction, obtained for
  $Rm=15$ and $S=2.3$.  Left: Radial velocity $U_r$ in the $(r,z)$
  plane. Curves are streamlines of the poloidal flow. Right: radial
  magnetic field component $B_r$(normalized by the applied field
  $B_Z^0$) with magnetic field lines contained in the $(r,z)$
  plane. For these boundary rotations [eq.~\eqref{Ome1}], MRI is
  characterized by a strong inflowing jet.}
\label{MRIpcolor}
\end{figure}

Figure \ref{MRIpcolor} shows the structure of the MRI mode obtained in
the saturated regime, for $Rm=15$ and $S=2.3$. Here, the structure of
the MagnetoRotational Instability mainly consists of two large
poloidal cells, as shown in the left panel. The
corresponding radial magnetic field is shown on the right.
At the endcaps, the fluid is strongly ejected
and generates an outflowing narrow jet. The recirculation takes the
form of a broad inflowing jet \cite{Liu08}. Consequently, we see that
the MRI modes resemble the hydrodynamic Ekman circulation obtained
when the endcaps corotate with the outer cylinder, except that the circulation is reversed.
In fact, MRI modes are relatively similar
to the classical Taylor vortices obtained in hydrodynamical
Taylor-Couette flow.

\begin{figure}
\centerline{
\epsfysize=65mm 
\epsffile{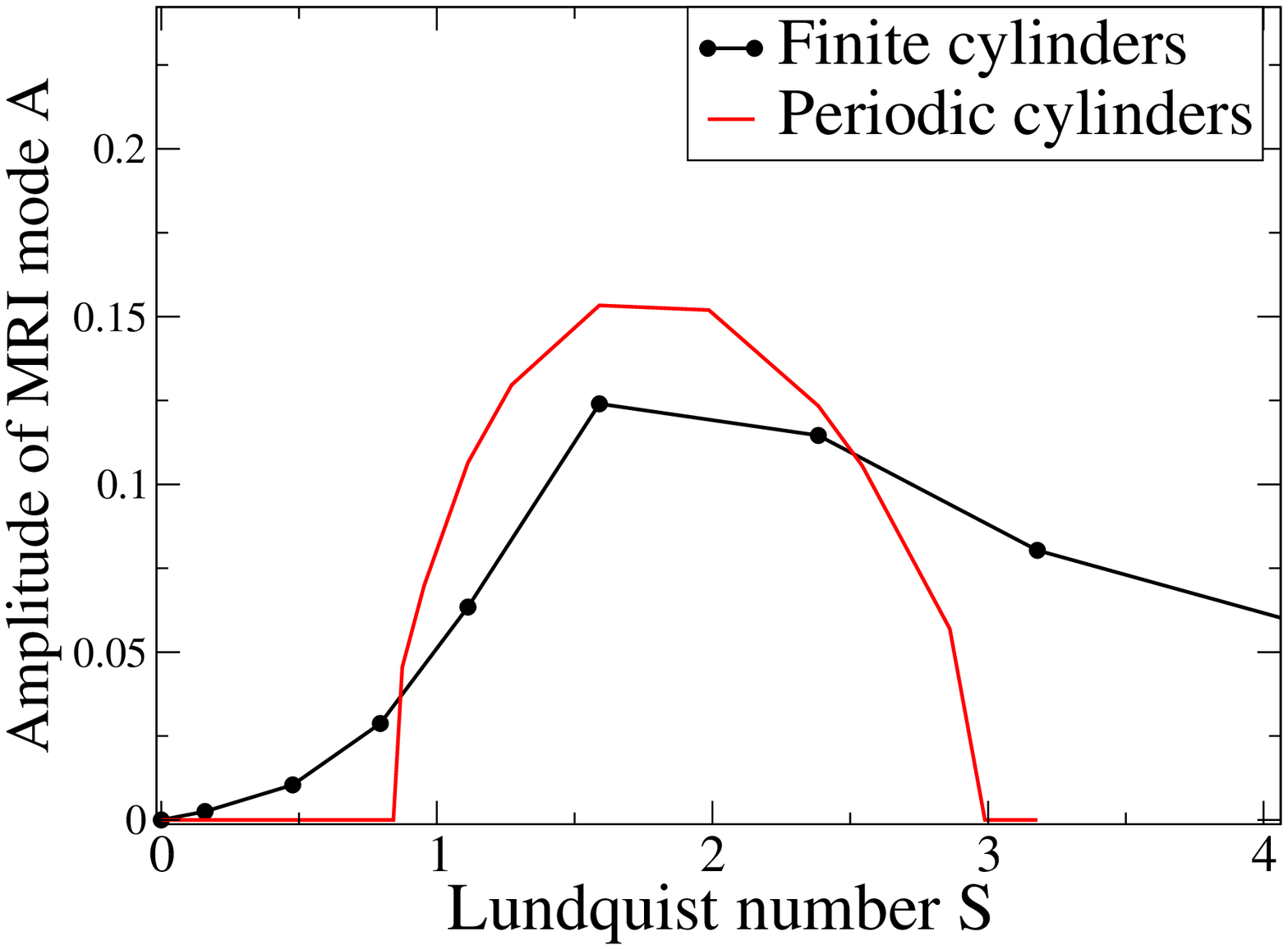} }
\centerline{
\epsfysize=65mm 
\epsffile{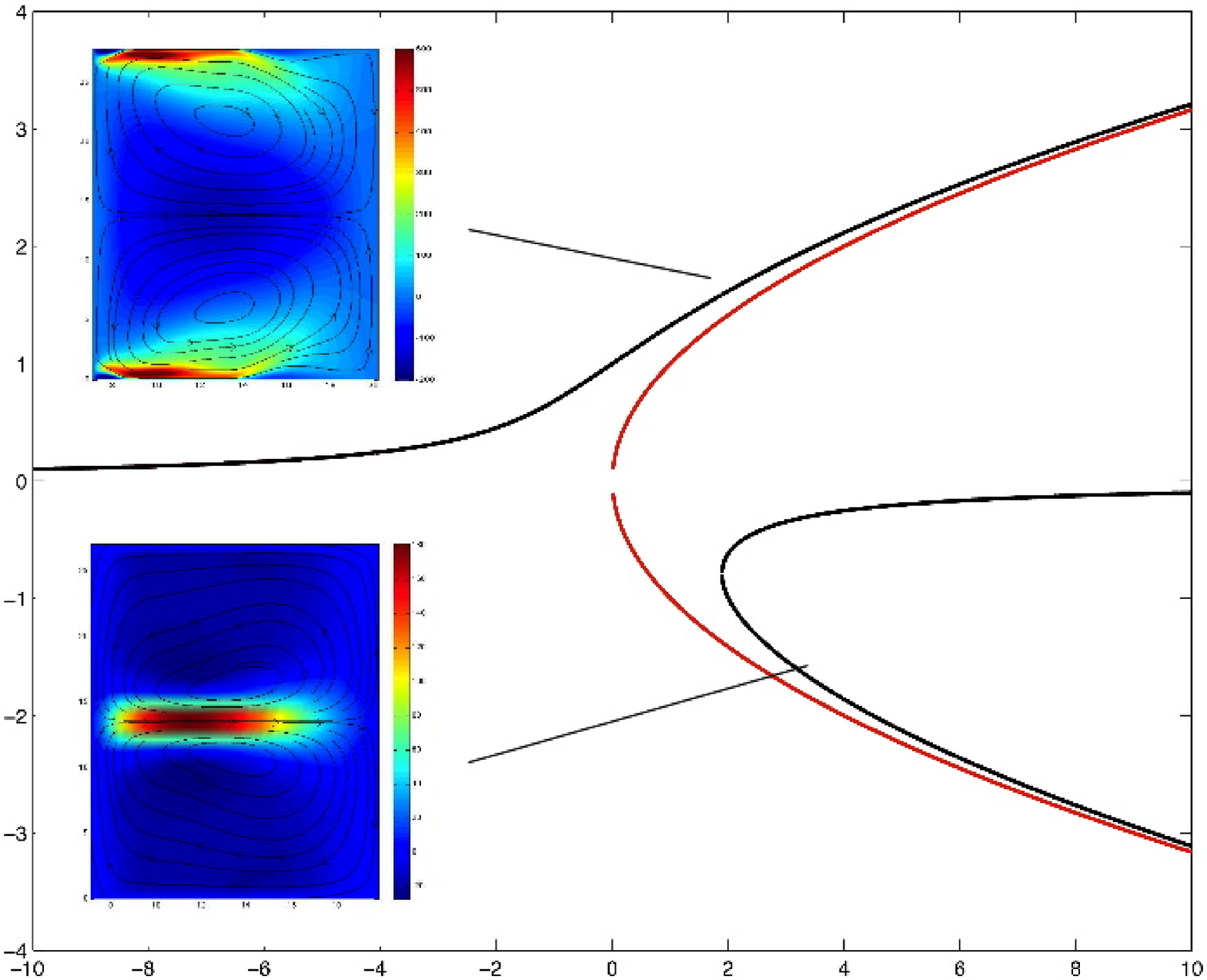} }
\caption{Top: Bifurcation diagram of the radial magnetic energy $A$ as
  a function of $S$ for $Rm=15$. Red curve correspond to periodic
  boundary conditions in $z$, whereas the black one corresponds to
  results obtained with no-slip rotating rings at the endcaps. In the
  latter case, the bifurcation is an imperfect supercritical pitchfork
  bifurcation due to the poloidal flow.  Bottom: Schematic
  representation of perfect (red) and imperfect (black) pitchfork
  bifurcations of the MRI. Each branch corresponds to a different
  structure of the MRI mode: insets show MRI mode obtained from
  rotation rates given by eq.(\ref{Ome1}) (top-inset) or
  eq.(\ref{Ome2}) (bottom-inset).}
\label{bifMRI}
\end{figure}

To follow the evolution of this large scale MRI mode, we compute the
global quantity $A=\sqrt{V^{-1}\int_V{B_r^2dV}}$, where $V$ is the
total volume between cylinders. Figure \ref{bifMRI} shows the
saturated value of $A$ as a function of the Lundquist number, for
$Rm=15$.  To understand how the MRI is generated in our finite
geometry, it is useful to compare this bifurcation diagram to results
obtained with periodic (or infinite) cylinders in the $z$
direction. In Figure~\ref{bifMRI}, the black curve thus indicates
results obtained with our no-slip rotating endcaps, whereas the red
curve shows the diagram obtained when vertically periodic boundary
conditions are used.\\

With periodic boundary conditions, the bifurcation of $A$ is
characterized by a well defined critical Lundquist number $S_c$ above
which MRI is observed ($S_c\sim 0.25$ for $Rm=15$). The magnetic field
follows a classical square root law $A\sim \pm\sqrt{S-S_c}$ slightly
above the MRI threshold $S_c$, and is zero below $S_c$.
In this case, the solution is relatively
similar to the one obtained with no-slip boundaries, except that the
axial position of the outflowing jet is made arbitrary by the periodic
boundaries, and the linear MRI modes are sinusoidal in the $z$
direction. In the linear regime, MRI modes correspond to the unstable
branch appearing from the coalescence of two complex conjugate
MagnetoCoriolis (MC) waves \cite{Nornberg10}. If these MC-waves are
linearly stable (as in our simulations), the non-linear transition
to MRI corresponds, at least close to the threshold $S_c$, to a
classical supercritical pitchfork bifurcation:
\begin{equation}
\dot{A}=\mu A-A^3
\label{perfect}
\end{equation}
where the dot denotes time derivative, and $\mu \propto S-S_c$ is a
control parameter. Each of the two supercritical branches of solutions
($A=\pm \sqrt{\mu}$) is a transposition of the other corresponding to
a phase-shift of one-half period, reflecting the arbitrary position of
the MRI jet. For instance, the upper branch will correspond to an MRI
mode with an outflow in the midplane, whereas the lower branch will
represent MRI mode with inflow in the midplane. A schematic view of
this supercritical bifurcation is shown in the lower panel of Fig.~\ref{bifMRI} by the red line.

However, no-slip vertical boundary conditions significantly change the
nature of the bifurcation: with finite cylinders, we see in figure
\ref{bifMRI},top (black curve) that the evolution of $A$ is more
gradual, and the definition of a critical onset $S_c$ is not well
defined. In a finite geometry, the magnetorotational instability
rather corresponds to an imperfect supercritical pitchfork
bifurcation:
\begin{equation}
\dot{A}=\mu A-A^3+h
\label{imperfect}
\end{equation}
Here, $h$ is a symmetry-breaking parameter reflecting the absence of
$z$-periodicity of the solutions, and forbidding the $A\rightarrow -A$
symmetry obtained in the case of periodic or infinite cylinders.
Figure \ref{bifMRI},bottom shows a schematic diagram of supercritical
pitchfork bifurcations, in perfect and imperfect cases. Physically,
this symmetry breaking means that for any value of the applied
magnetic field, there is always a non-zero solution corresponding to
the small poloidal recirculation created by the endcaps (see Figure~\ref{Couette},bottom-left).
Therefore, with realistic boundaries, {\it the
magnetorotational instability continuously arises from this residual
magnetized return flow, so that the axial position of the MRI outflowing
jet is no longer arbitrary}. This can be regarded as an analogue of the
transition encountered in finite Taylor-Couette flow, in which Taylor
vortices gradually emerges from the Ekman flow located at the endcaps
\cite{Benjamin81}. Figure \ref{bifMRI}-top also shows that for strong
applied field, the magnetic tension suppresses MRI modes. It is
interesting to note that again, this restabilization is much more
gradual in finite geometry than in the periodic case.\\

The modified nature of the bifurcation in the presence of realistic
boundaries has important implications for experimental observations of the
instability, because it strongly complicates the
definition of an onset for the MRI. Since a recirculation is always
generated by the boundaries, it will be difficult to distinguish an MRI
mode from a simple magnetized meridional return flow, at least close
to the instability threshold.
If the generation of the magnetorotational instability in laboratory
experiments corresponds to a strongly imperfect pitchfork bifurcation,
as suggested by our numerical results, one can expect the following:\\

- First, the MRI will appear with the imperfect scaling obtained after
integration of equation (\ref{imperfect}) instead of a classical
square root law. On the one hand, this means that the transition to
instability in bounded systems will not be as clear-cut as formerly expected,
especially in experiments where only small $Rm$ can be
obtained. On the other hand, the gradual transition to MRI means that
some of the typical features of the instability can be observed below
the expected onset of the MRI.\\

-Different structures for the MRI are obtained depending on the
hydrodynamical configuration. This is directly related to the
imperfect nature of the bifurcation: the residual flow due to the
boundaries strongly affects the geometry of the unstable mode. For
instance, the mode shown in Figure \ref{MRIpcolor} has been obtained
from the hydrodynamical state given by eq.(\ref{Ome1}) and shown in
Figure \ref{Couette}. Without magnetic field, a weak outflow near the
endcaps is produced, and the MRI mode generated in this case consists
of poloidal vortices with the same outflowing jet near the endcaps. In
other words, the hydrodynamical flow favors the upper branch of the
pitchfork bifurcation, as in Figure \ref{bifMRI}-bottom. A
different hydrodynamical configuration would lead to a different MRI
mode. Consider for instance the following rotation rates:
\begin{equation}
\Omega_1=1. \ , \hspace{10pt}
\Omega_2= \  \hspace{3pt}
\Omega_3= \  \hspace{3pt}
\Omega_4=0.13 \ 
\label{Ome2}
\end{equation}
Here, the endcap rings are rigidly rotating with the outer cylinder.
The unmagnetized flow is characterized by a strong inflow near
endcaps, and the outflow is now localized in the midplane. In the presence
of a magnetic field, MRI modes show a similar geometry, with a strong
outflow in the midplane. This corresponds to change $h$ to $-h$ in the
equation (\ref{imperfect}), thus selecting the lower branch mode $-A$
instead of $A$ (this MRI mode is shown in the lower inset of Figure
\ref{bifMRI}).\\

-Finally, Figure \ref{bifMRI}-bottom illustrates that in the case of an
imperfect bifurcation, one of the branches is replaced by a saddle-node
bifurcation of two modes, one stable and the other unstable,
predicting the possible co-existence of two MRI modes with different
geometries. Note that the transition to this new mode would be
strongly subcritical. This is the equivalent of the well known ``{\it
  anomalous modes}'' observed in hydrodynamical Taylor-Couette
flows. Due to its subcritical nature, this mode has not been observed
in our simulations, the lower inset in Figure \ref{bifMRI}-bottom
corresponding to a simulation with rotation rates given by
eq. (\ref{Ome2}), when the {\it lower} branch is continously connected
to the Ekman flow.\\

\subsection{Saturation of the MRI}

\begin{figure}
\centerline{
\epsfysize=55mm 
\epsffile{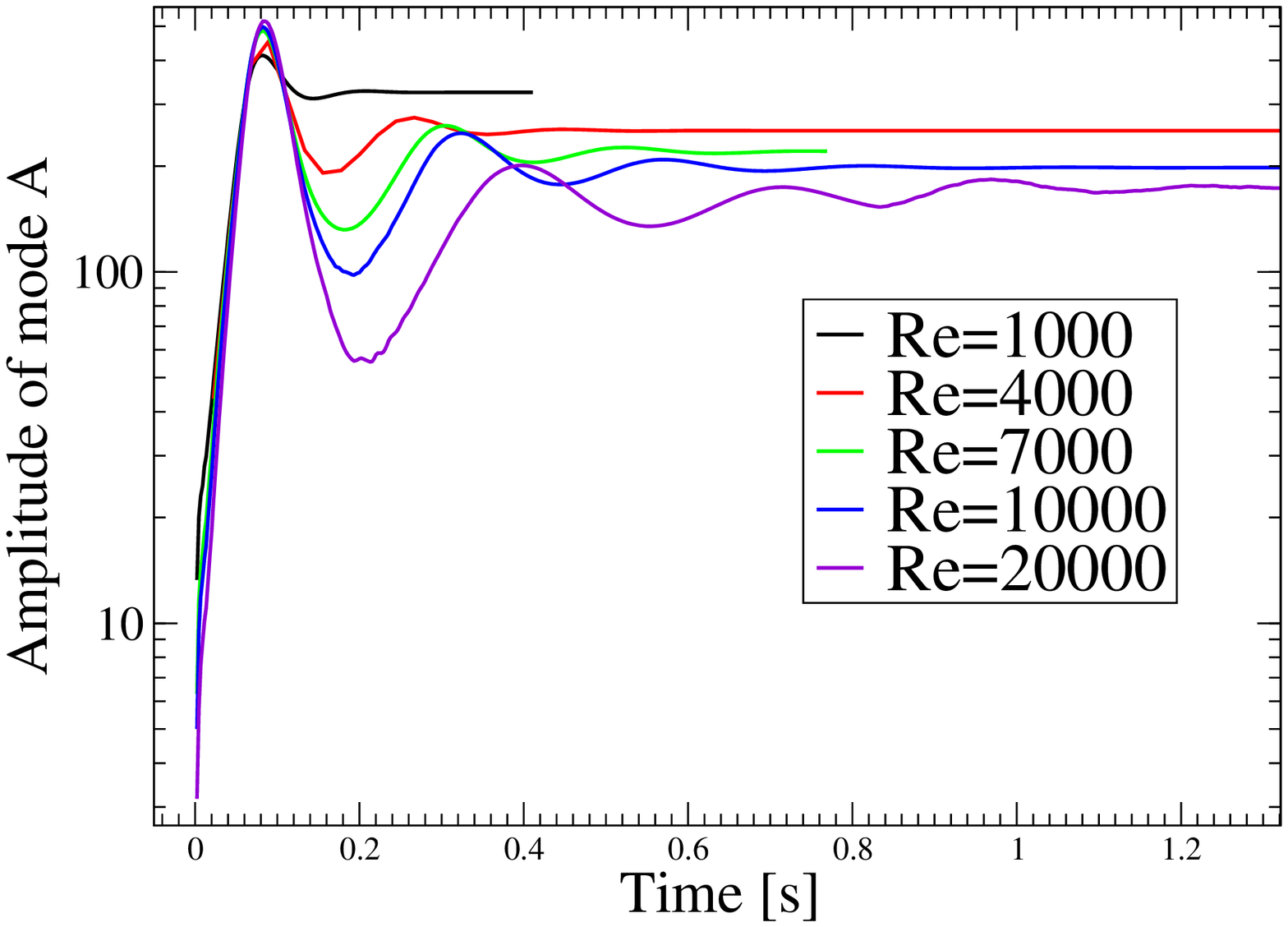}} 
\centerline{
\epsfysize=55mm 
\epsffile{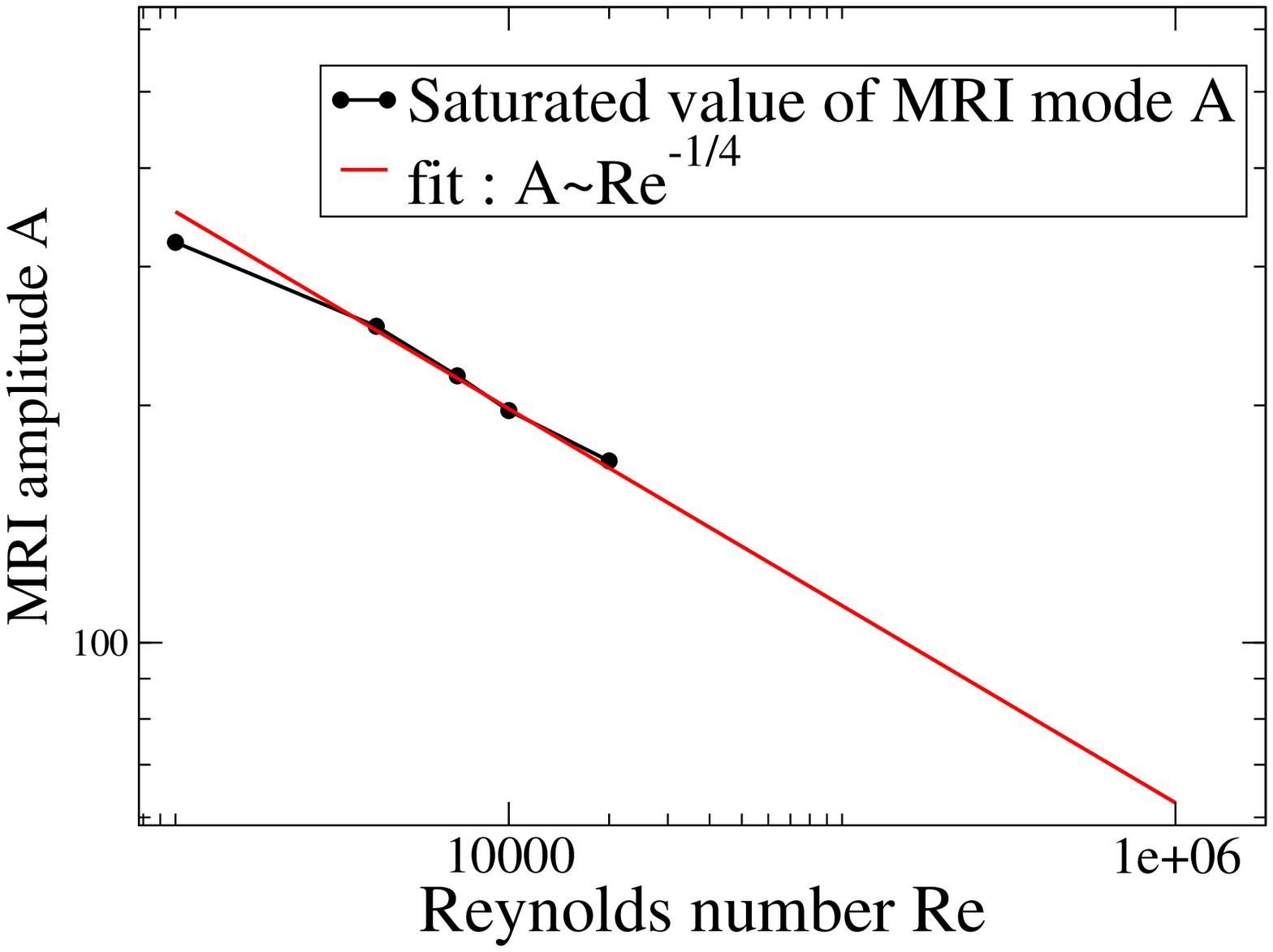} }
\caption{ Effect of the hydrodynamical Reynolds number Re on the
  growth of MRI modes. The magnetic Reynolds number and the Lundquist
  number are kept fixed to $Rm=30$ and $S=4.5$. Top: Time evolution of
  the radial magnetic field density $A$ as a function of time, for
  different values of $Re$. Linear growth rates are independent of
  $Re$ but the saturation level strongly depends on $Re$. Bottom:
  evolution of the saturated value of $A$ as a function of $Re$. The
  saturation value of the MRI strongly decreases with $Re$. Red line
  indicates a fit corresponding to the scaling $A \sim Re^{-1/4}$}
\label{MRIRe}
\end{figure}

In all the simulations reported here, most of the parameters have
values comparable to those of laboratory experiments. For instance, in the
Princeton experiment, the magnetic Reynolds number $Rm$ can be varied
from $0$ to $15$, whereas the Lundquist number $S$ is between $0$ and
$3$. The exception is the hydrodynamical Reynolds number, which can exceed
$10^6$ in the experiments, whereas our simulations are limited by numerical resolution
to $Re<2\times 10^4$. It is thus
important to understand how our numerical results scale to a
system with $Re\sim 10^6$. Figure~\ref{MRIRe} shows the evolution of
the saturation of the MRI as a function of the Reynolds number.

Figure \ref{MRIRe}-top shows the time evolution of the radial magnetic
field density for different Reynolds numbers, ranging from $Re=10^3$
to $Re=2\times10^4$. In the linear phase, all the growth rates are the
same, illustrating the relevance of the linear inviscid theory and the
weak role played by the Reynolds number. On the contrary, the
saturation level depends on the Reynolds number: as $Re$ increases,
the instability saturates to smaller values. It seems that the
saturation of the MRI in our simulations scales like $A \sim
Re^{-\beta}$, with $\beta=1/4$. This is a dramatic illustration of the
challenges of observing MRI experimentally, where $Re$ must be several
millions in order to achieve $Rm>1$ because of the properties of
liquid metals. However, a naive extrapolation of our numerical results
suggests that the magnetorotational instability in the Princeton
experiment should saturate at a value only one order of magnitude
smaller than in the simulations reported here, and should be
measurable. \\

This scaling with $Re$ is directly related to the nature of the
non-linear saturation mechanisms involved in Taylor-Couette flows.
Indeed, it has been suggested by Knobloch et al \cite{Knobloch05} that
in such Taylor-Couette devices, saturation mainly occurs by
modification of the mean azimuthal profile: the shear in the flow is
reduced, and is ultimately controlled by viscous coupling to the
boundaries.  Note that a simple dimensionless analysis shows that the
exponent $\beta$ of our scaling corresponds to a balance between the
Ekman pumping ($\sim Re^{-1/2}$) and the Maxwell stress tensor ($\sim
B^2$), in agreement with this scenario. However, it is important to
note that the saturation of the MRI is expected to be strongly
different in discs, where shear is fixed and boundaries may be
stress-free.

\section{Non-axisymmetric instabilities}

\begin{figure}
\centerline{
\epsfysize=50mm 
\epsffile{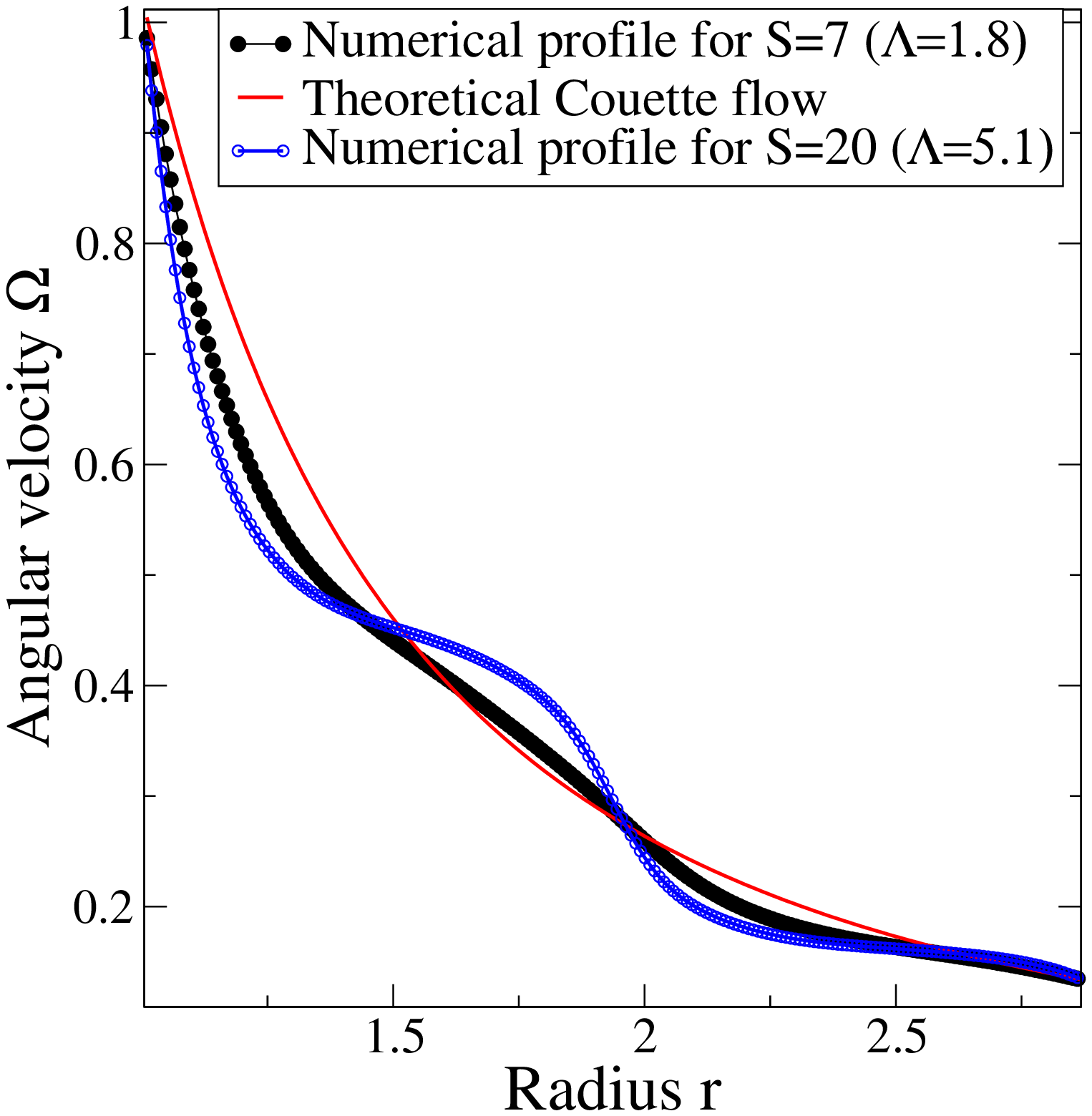} 
\epsfysize=50mm 
\epsffile{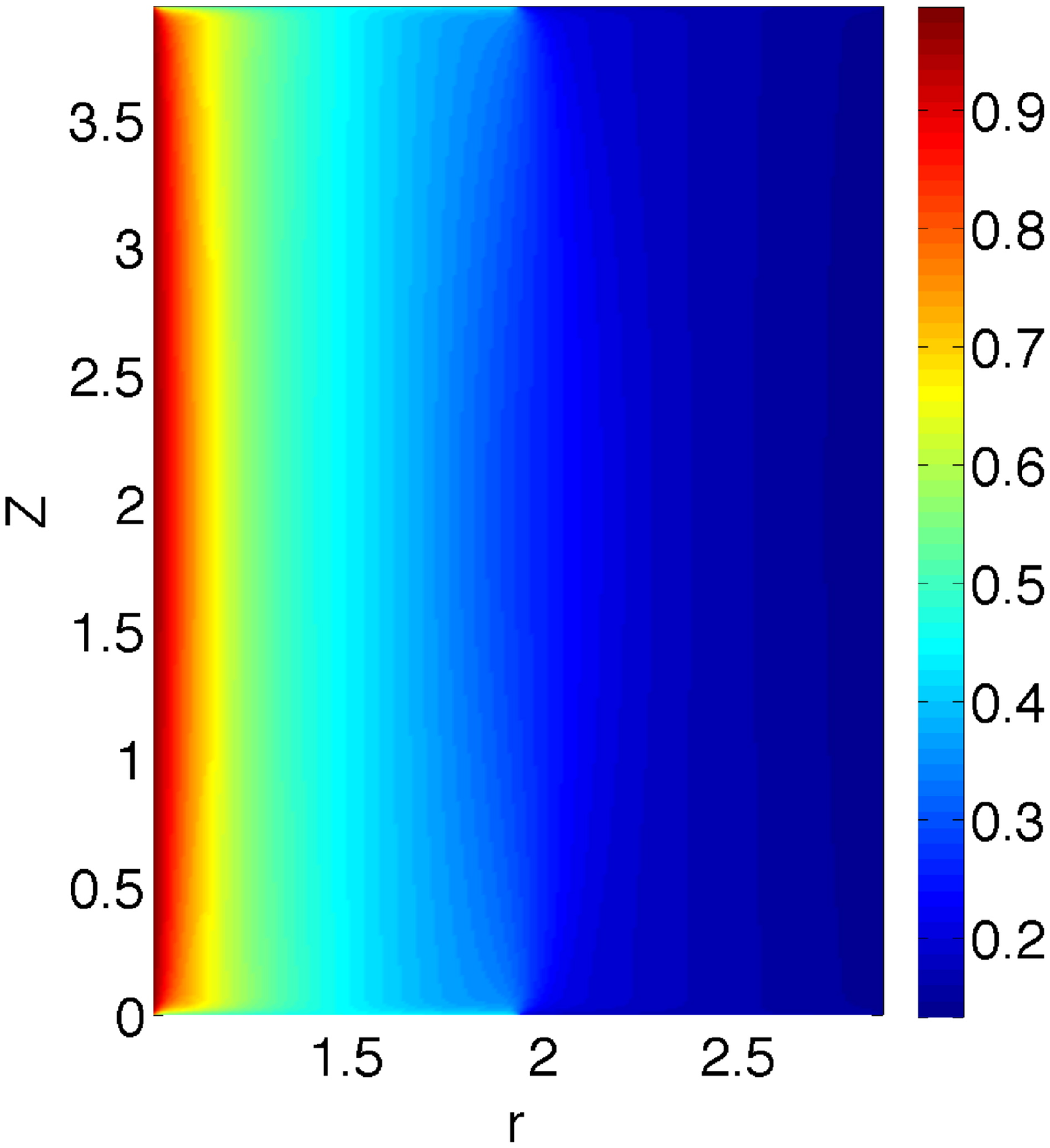} }
\centerline{
\epsfysize=50mm 
\epsffile{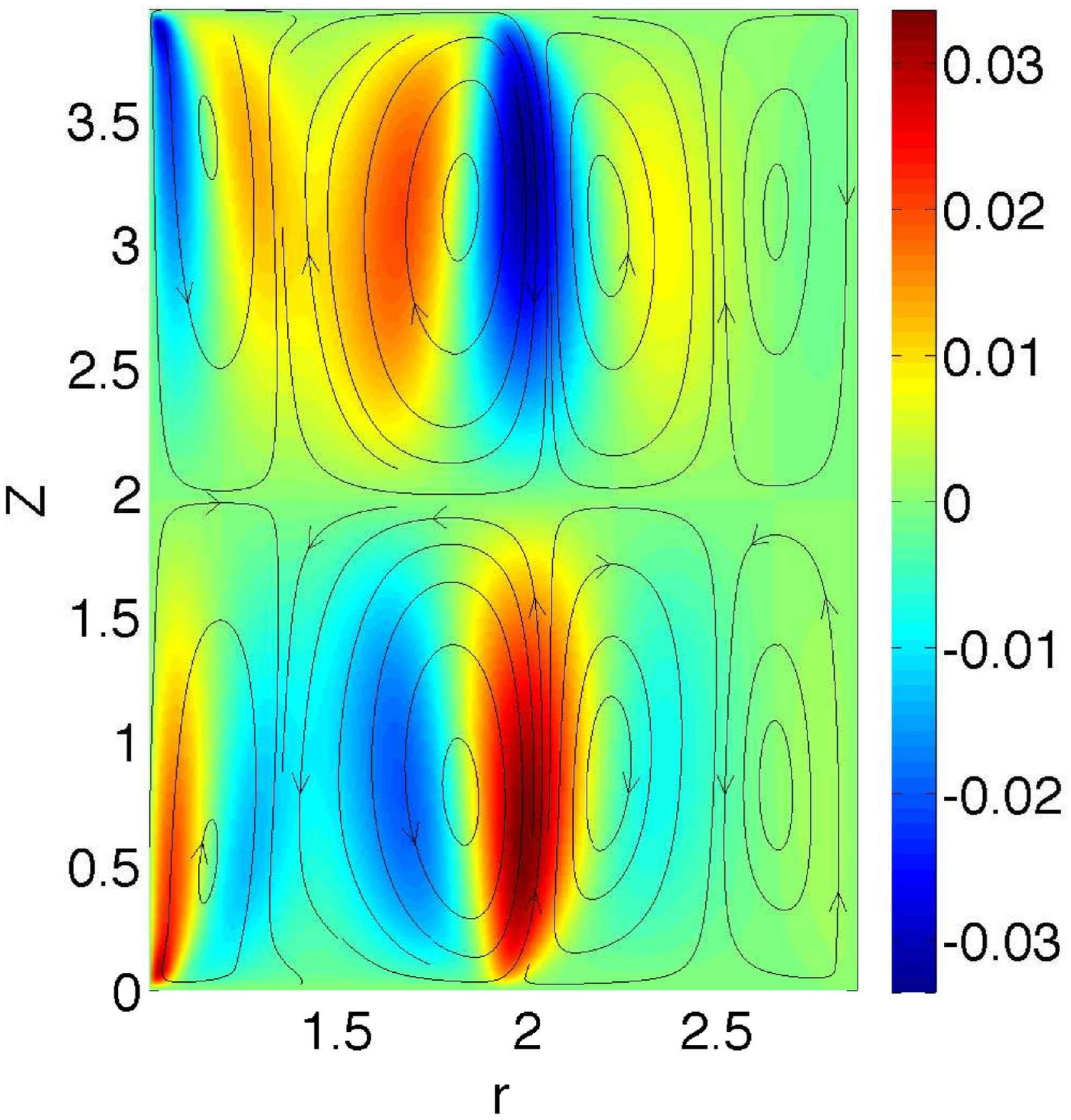} 
\epsfysize=50mm 
\epsffile{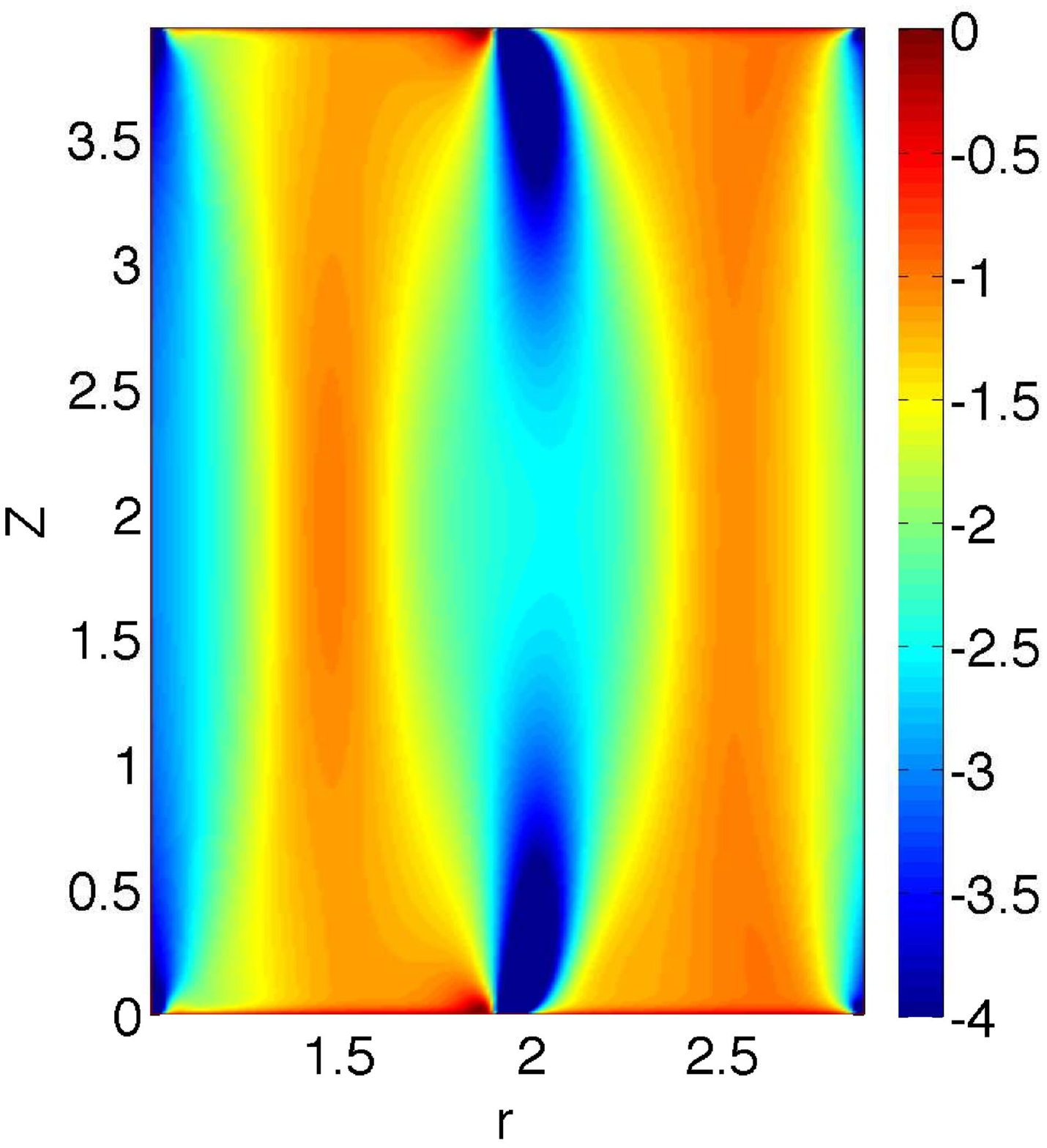} }

\caption{Taylor-Couette flow when a strong uniform magnetic field
  $B_z$ is applied in the axial direction $z$, obtained for
  $(\Omega_1,\Omega_3,\Omega_4,\Omega_2)=(400,180,65,53)$, $Rm=15$,
  $S=7$.  Top-left: profile of $\Omega$ at $z=h/4$. Top-right: Angular
  velocity $\Omega$ in $(r,z)$ plane. We also show the profile
  obtained for $S=20$.  Bottom-left: $U_z$ in the $(r,z)$ plane
  (streamlines indicate meridional flow. Bottom-right: Shear $q$ in
  the $(r,z)$ plane. The magnetic field tends to homogenize the flow
  in the $z$-direction. The jump of velocity  between rings at endcaps
  extends into the bulk of the flow, generating a free shear layer at
  $r=(r_1+r_2)/2$.  The $8$ poloidal cells are re-enforced and a
  vertical jet is created. Note that this corresponds to the flow
  before the saturation of the non-axisymmetric instability.}
\label{CouetteBz}
\end{figure}

Let us now consider strongly magnetized
situations. Figure~\ref{CouetteBz} shows the same as
Figure~\ref{Couette}, but for an applied field such that $S=7$. In all
the simulations reported here, the magnetic field is applied at the
beginning of the simulation when the flow is at rest, letting the flow
dynamically adapts as time evolves.  At these large fields, we first
note that the angular velocity profile is significantly different from
the ideal Couette profile. In particular, the profile is steeper close
to the inner cylinder and around $r=(r_1+r_2)/2$. The distribution of
the shear parameter $q$ (Fig. \ref{CouetteBz}c) shows that the
Stewartson shear layer (characterized by $q>-3.5$), confined to the
endcaps in the purely hydrodynamical case, now penetrates deep inside
the bulk of the flow \cite{Liu08b}: the applied magnetic field, by
uniformizing the flow in the $z$-direction, reinforces and extends the
shear between rings. This leads to the generation of a new free shear
layer, different from the Stewartson layer. In addition, note that the
poloidal flow is strongly affected by the applied field. The eight
poloidal cells generated in the non-magnetic case are reinforced, and
aligned in the $z$-direction.  A strong axial jet develops at
$r=(r_1+r_2)/2$, where the flow is ejected from the ring separation to
the midplane.\\

A similar magnetized free shear layer has been extensively described
in spherical geometry. The shear layer appears when a
strong magnetic field is applied to a spherical Couette flow, the
magnetic tension coupling fluid elements together along the direction
of magnetic field lines.  In the case of an axial magnetic field, this
creates a particular surface $\Sigma$ located on the tangent cylinder,
the cylindrical surface tangent to the inner sphere. Fluid inside
$\Sigma$, coupled by the magnetic field only to the inner sphere,
co-rotates with it, whereas fluid outside $\Sigma$ (coupled to both
spheres) rotates at an intermediate velocity. The jump of velocity on
the surface $\Sigma$ therefore results in a free shear layer,
sometimes called Shercliff layer \cite{Shercliff53}. The thickness
$\gamma$ of this MHD free shear layer is expected to vary like $\gamma
\sim r/\sqrt{M}$, where $M$ is the Hartmann number. In our case, the
Shercliff layer occurs on the cylindrical surface of radius
$r=(r_1+r_2)/2$, which separates regions of fluid coupled to the inner
rings from fluid coupled to the outer rings.

In spherical geometry, it has been shown that this flow configuration
is unstable, and either the Shercliff layer or its associated poloidal
flow leads to the generation of non-axisymmetric modes. In a recent
work \cite{Gissinger11}, it has been suggested that these modes,
rather than MRI, were
responsible for the non-axisymmetric oscillations observed in the
Maryland experiment.
Interestingly, similar non-axisymmetric instabilities
are observed in our cylindrical configuration, despite the difference
in the geometry and the forcing. The red curve in Figure~\ref{curve}
is the locus of marginal stability for these non-axisymmetric
modes; modes are unstable to the right (larger $S$ for given $Rm$)
of the curve. The critical Reynolds number
seems to follow the scaling $Rm_c\sim S^2$. These non-axisymmetric
modes are also restabilized for strong values of the applied field,
but we do not have enough points to determine the corresponding
scaling.  

The marginal-stability curve $Rm_c\sim S^2$ corresponds in fact to the
region of the parameter space where the Elsasser number,
$\Lambda\equiv B^2/(\mu\rho\eta\Omega)$, is equal to unity. From
Figure~\ref{curve}, it can be seen that our shear layer instability
extends to small magnetic Reynolds
number ($Rm<1$). Apparently, the shear layer instability reported here
is inductionless, meaning that the time derivative of the magnetic field can be
neglected in the induction equation \eqref{ind}.
This is a crucial difference from the standard magnetorotational instability,
for which induction is essential ($Rm\gtrsim1$).
The fluid Reynolds number, $Re$,
is more relevant to the present instability than $Rm$.
This inductionless property is also possessed by the so-called
Helical MRI, which has been understood as an inertial wave weakly
destabilized by the magnetic field
\cite{Kirilov10}, \cite{Liu06b}.  The inductionless nature of the
shear-layer instability is easy to understand physically: although a
strong magnetic field is necessary to generate the layer, the
destabilization of this layer is triggered by a hydrodynamical
shear instability, similar to Kelvin-Helmoltz, whose form is nearly constant
along the field lines and therefore scarcely affected by magnetic tension.
The instability is suppressed if the
difference between the rotation rate of the two rings is too small.\\

\begin{figure}
\centerline{
\epsfysize=70mm 
\epsffile{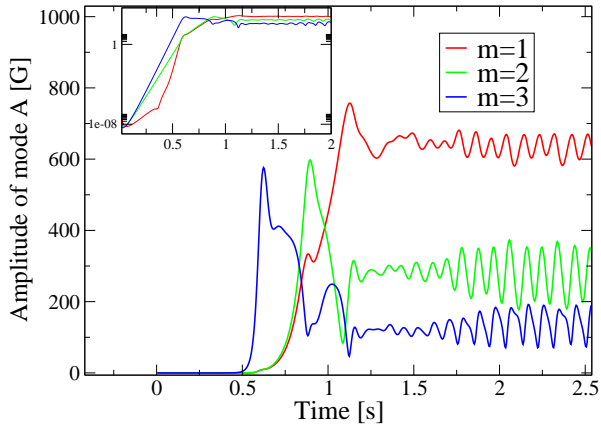}}
\caption{Time evolution of Fourier components of the azimuthal
  velocity. The shear layer is destabilized to non-axisymmetric
  modes via a Kelvin-Helmoltz type instability. In the linear phase,
  the $m=3$ azimuthal harmonic is the most unstable;  the saturated state is
  dominated by $m=1$. Inset: the same evolution as a semilog plot.}
\label{Sher_time}
\end{figure}

Figure~\ref{Sher_time} shows the time evolution of azimuthal Fourier
modes for $Rm=15$ and $S=11$. The initially axisymmetric MHD shear
layer is destabilized to several azimuthal wavenumbers. In the linear
phase, the $m=3$ mode clearly dominates. As the instability saturates,
there is a cascade of energy towards lower azimuthal wavenumbers, and
the nonlinear saturated state is strongly dominated by an oscillating
$m=1$ mode. In the saturation phase, the thickness of the shear layer
is increased, indicating that the instability saturates by modifying
the shear profile.

\begin{figure}
\centerline{
\epsfysize=50mm 
\epsffile{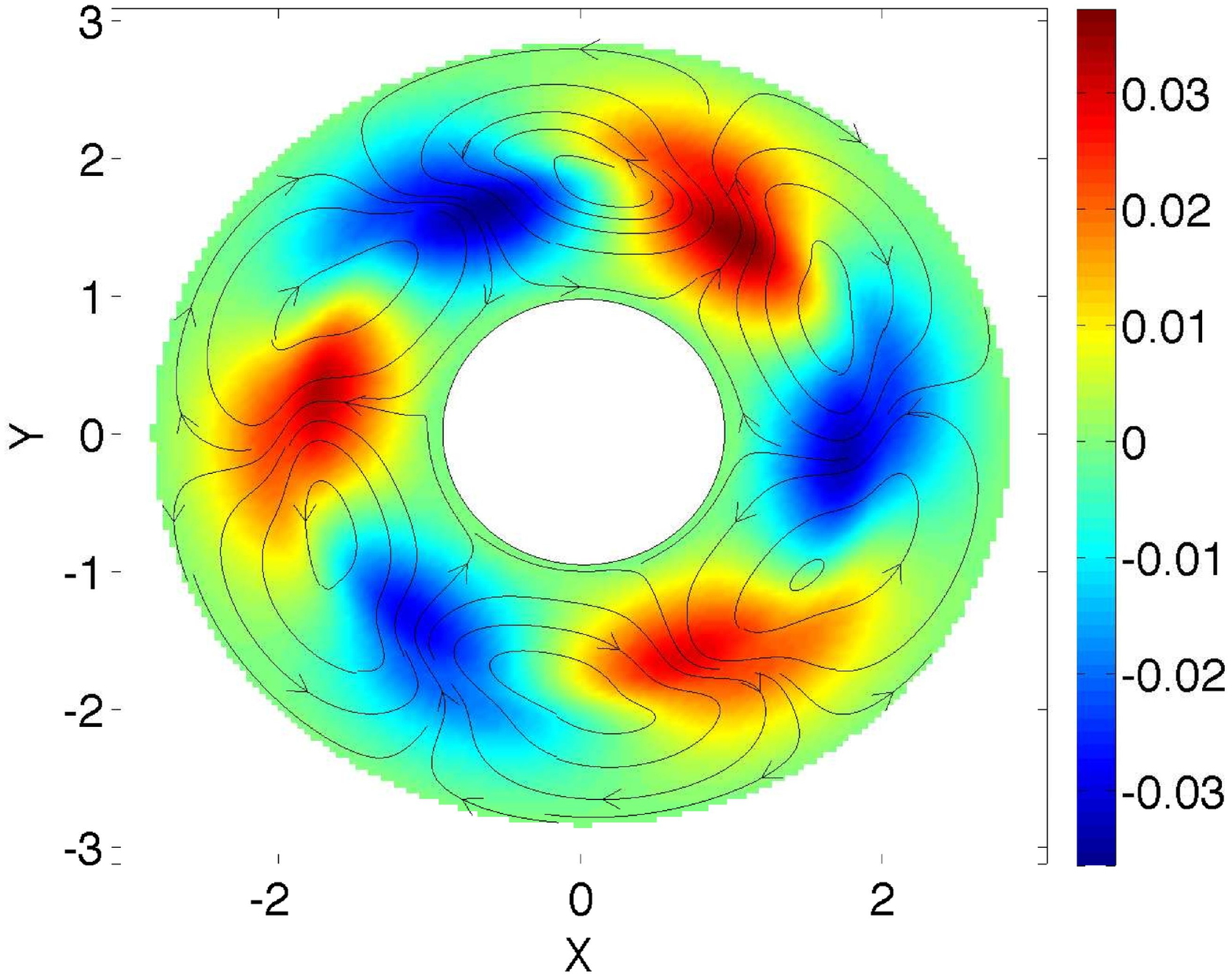} }
\centerline{
\epsfysize=50mm 
\epsffile{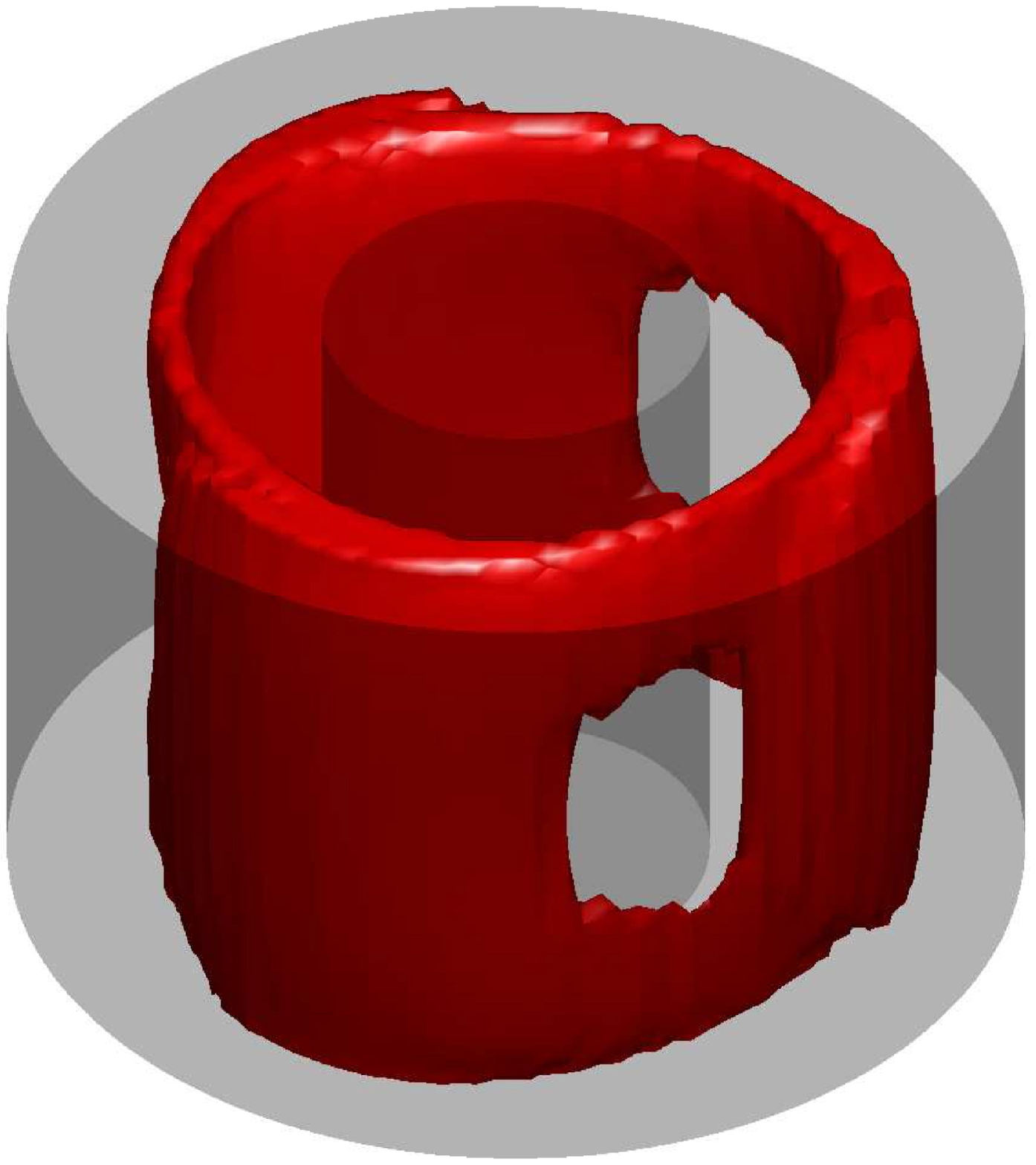} }
\caption{ Structure of the non-axisymmetric instability for $Rm=15$,
  $S=11$. Top: Structure in the $r,\phi$ plane, for $z=H/4$.
  Non-axisymmetric component of the radial velocity $u_r$ in the
  equatorial plane during the linear phase. Bottom: $3D$ isosurface of
  the shear $q$. The mode consists of non-axisymmetric axial columnar
  modes, i.e.  vortices in the $r,\phi$ plane (the axial component is
  much weaker) and mainly independent of the $z$-direction. Isosurface
  of the shear shows that these modes are essentially Kelvin-Helmoltz
  destabilization of the central magnetized shear layer.}
\label{Sher_struc}
\end{figure}

Figure~\ref{Sher_struc} shows the structure of these non-axisymmetric
instabilities. In the linear phase, the structure is very similar to
Kelvin-Helmoltz modes of the free shear layer, taking the form of
columnar vortices transverse to the $(r,\phi)$ plane, independent
of the axial direction and therefore symmetrical with respect to the
equatorial plane. Note that the vortices spiral slightly in
the prograde direction, as expected from a Kelvin-Helmoltz instability
in cylindrical geometry transporting the angular momentum
outward. Figure~\ref{Sher_struc}-top shows an isosurface of $25\%$
of the total shear $q$ in the flow. This illustrates how the
rotational symmetry of the free shear layer is broken in the $\phi$
direction, the initially axisymmetric layer rolling up in a series of
horizontal vortices by a mechanism similar to Kelvin-Helmoltz
instabilities.\\

The structure of the magnetized shear layer strongly depends on the
parameters. For a fixed value of the global rotation, the thickness
$l$ of this shear layer scales with the Hartmann number as
$l\sim r/\sqrt{M}$, a scaling similar to what is observed in
spherical geometry. On the other hand,  if the applied field is decreased
after this shear layer has formed, the vertical extent of the
shear layer is reduced, and eventually reaches a critical length where
the Kelvin-Helmoltz instability does not occur. In our simulations, it
appears that the vertical extent of the shear layer scales as
$h\sim rM/\sqrt{Re}=r\sqrt{\Lambda}$. The condition $h\sim r$
thus gives the $\Lambda=1$, destabilization condition observed in our
simulations: {\it the non-axisymmetric modes develop by
  Kelvin-Helmoltz destabilization of the shear layer, provided that
  the layer is sufficiently extended in the vertical direction
  ($\Lambda>1$)}.

\section{Comparison with laboratory experiments}

Finally, we compare our numerical simulations to results obtained in
the Princeton MRI experiment, where a similar setup is used (see
\cite{Roach11} for more details on the experimental results mentioned
here).  Because of the very small magnetic Prandtl number of liquid
metals ($Pm=Rm/Re\lesssim 10^{-5}$), it is very challenging to achieve
fluid Reynolds numbers large enough to observe MRI in liquid-metal
experiments. In Figure~\ref{curve}, the region of parameter space
accessible to the Princeton experiment is indicated by the dashed
square.

Our numerical simulations indicate that the magnetorotational instability should in theory
be observable in the Princeton MRI experiment between $Rm=10$ and $Rm=15$; the latter
corresponds to the maximum designed rotation rotation speed. Up to now, the maximum
magnetic Reynolds number reached in the Princeton experiment is $Rm=9$, i.e. just below
threshold.  Our simulations suggest that the MRI should take the form of a gradual
amplification of the residual Ekman flow. Since the experiment is still operating very
close to the expected onset of the MRI, it is necessary to operate at higher $Rm$ to see
MRI clearly.  A further difficulty is that the saturation level is expected to be smaller
in the experiment that in the simulations because of the $Re^{-1/4}$ scaling discussed
above (Fig.~\ref{MRIRe}).

\begin{figure}
\centerline{
\epsfysize=48mm 
\epsffile{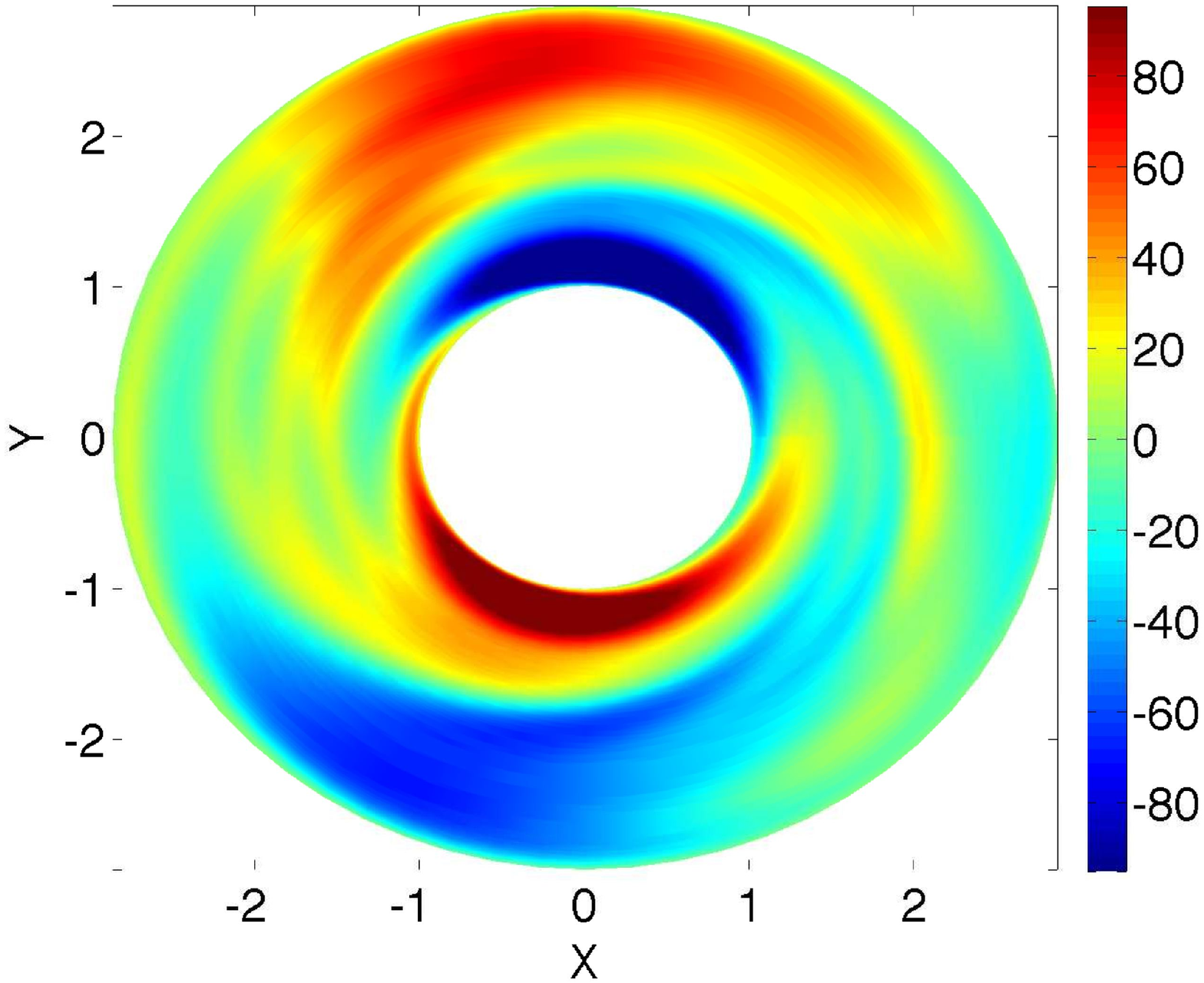} }
\centerline{
\epsfysize=48mm 
\epsffile{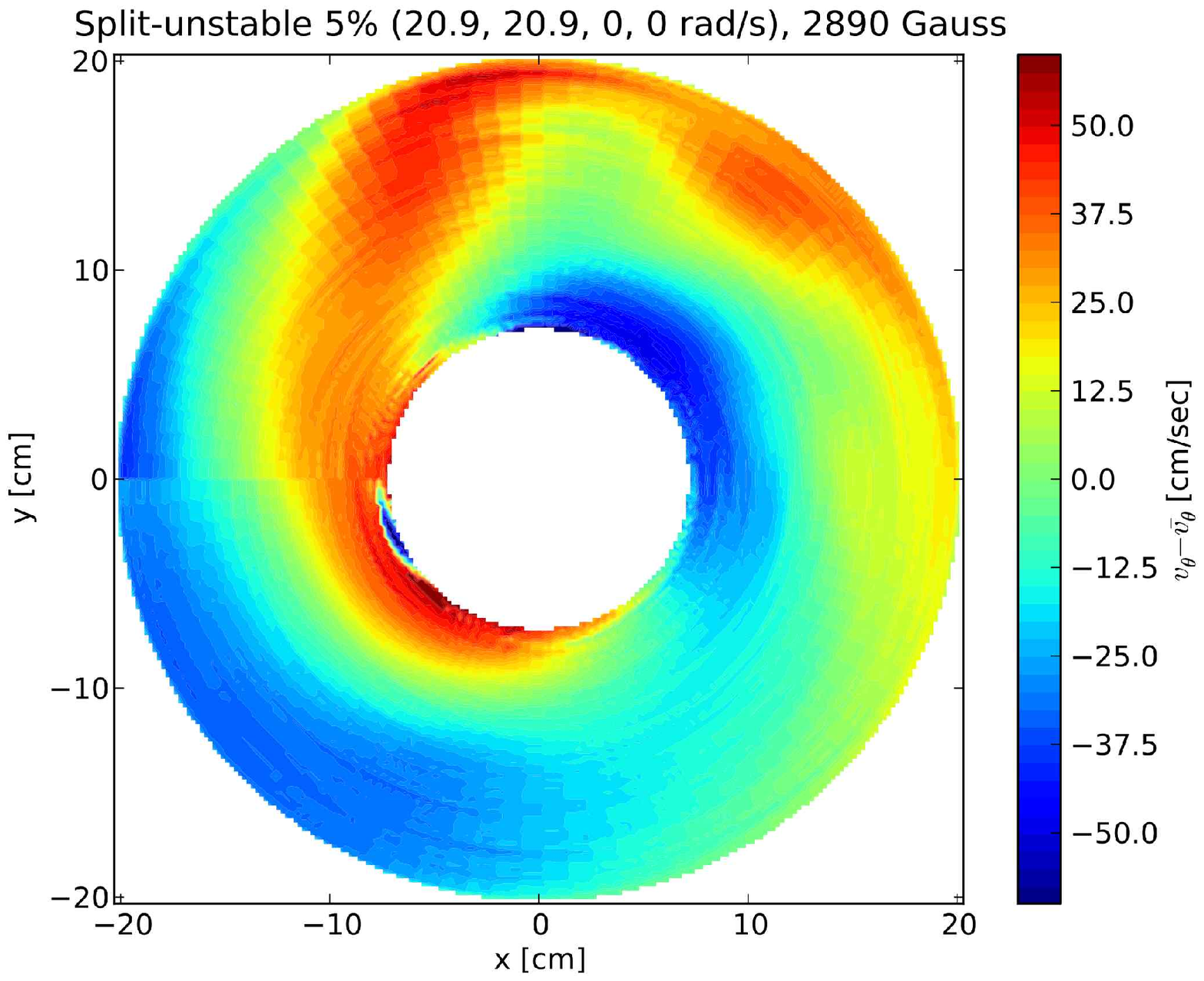}}
\caption{Comparison between numerical simulation (top) and ultrasound
  measurements \cite{Roach11} from the Princeton experiment
  (bottom). In both cases, inner rings corotate with the inner cylinder
($\Omega_1=\Omega_3$), outer rings and cylinder are at rest ($\Omega_2=0=\Omega_4$),
$Re=1450$, and $\Lambda=1.6$. Simulation has $Rm=0.02$, and experiment, $Rm=0.002$.
Figures show the non-axisymmetric component of the
  azimuthal velocity $U_\phi$ in the midplane, when the Shercliff 
  layer is unstable. A very good agreement is obtained: in both cases,
  the saturated state consists of an $m=1$ spiral vortices rotating in
  the azimuthal direction. In the experiment, $U_\phi$ is obtained
  from two ultrasound probes measuring the radial profile of the
  velocity at two different azimuthal positions. }
\label{num_exp}
\end{figure}

By contrast, the shear layer instability is inductionless and should
be observed at much smaller $Rm$. Figure~\ref{num_exp} compares the
numerical and experimental results for the horizontal structure of the
mode.  Non-axisymmetric modes have indeed been observed at very low
rotation rates in the experiment, and these modes present several
features identical to the instabilities reported here:\\

$\bullet$ First, the generation of non-axisymmetric modes in the experiment
follows the marginal stability curve $\Lambda_c=1$, in agreement with
our numerical prediction. Therefore, the experimental instabilities
are also inductionless: they
have been observed for rotation rates of the inner cylinder as small
as $\Omega_1=0.05\,{\rm rad\,s^{-1}}$ ($Rm\ll 1$).\\

$\bullet$ The structure of the modes is very similar. Figure~\ref{num_exp}
compares the geometry of the modes in the ($r,\phi$) plane obtained in
the experiment (bottom) and our simulations (top), for a case in which the
dimensionless experimental parameters match the numerical ones except in
magnetic Reynolds number.  The similarity of the instability across two orders
of magnitude in $Rm$ confirms its inductionless nature.
In both cases,
non-axisymmetric modes consist of equatorially symmetric vortices in
the $(r,z)$ plane taking the form of an $m=1$ spiral mode sheared in
the azimuthal direction.  Note however that in the simulation, close
to the instability onset, the modes spiral retrogradely in the
midplane, but tend to be more prograde close to the rings (in
agreement with the expected structure of a Kelvin-Helmoltz
instability). This effect has not been observed in the experiment,
where the spiral seems to be retrograde even relatively close to the
endcaps. This could be due to the effect of the mean shear flow, or
the secondary excitation of typical waves of the system, such as
magneto-coriolis waves \cite{Nornberg10}. A prograde spiral is however
expected sufficiently close to the endcaps, since it corresponds to an
outward transport of the angular momentum by the Kelvin-Helmoltz
instability. \\

$\bullet$ The time evolution is similar. In both numerical simulations and
experiment, the saturated state is an $m=1$ mode rotating at a
constant speed in the equatorial plane. 
Before reaching this $m=1$ mode, the experimental flow exhibits transient
oscillations with different azimuthal symmetry, namely an $m=3$
followed by an $m=2$ mode. This is similar to the behavior of our
numerical model, reported in Figure~\ref{Sher_time}.\\

$\bullet$ In the experiment, non-axisymmetric modes are observed only
when the difference of angular velocity between the inner ring and the
outer ring is sufficiently large, and global rotation can stabilize
the flow. These two features are also observed in our numerical
simulations.

These several similarities strongly suggest that the non-axisymmetric
modes observed in the Princeton MRI experiment are of the same nature
as the ones reported in our numerical simulations, namely
Kelvin-Helmoltz-type destabilization of a free shear layer created by
the conjugate action of the axial magnetic field and the jump of
angular velocity between the rotating rings at the endcaps. Note
however that the magnetic boundary conditions at the endcaps are not
exactly the same: our simulations use the idealized 'pseudo-vaccum'
boundary conditions given by eq.(\ref{fero}), whereas the endwalls are
insulating in the Princeton experiment. Since insulating endwalls do
not couple with magnetic field, the role of the electrical currents
close to the endcaps in the establishment of the layer could be
slightly different between experiments and numerics.\\

Interestingly, similar instabilities occur in spherical geometry, and
could be involved in the oscillations observed in the Maryland
experiment. More recently, similar rotating rings have also been used
in the PROMISE experiment in order to obtain clearer observation of
the Helical MRI, and non-axisymmetric $m=1$ modes have been
reported. It would be interesting to see how these oscillations
compare to the non-axisymmetric modes reported here.

\section{Conclusion}

In this article, we have reported three-dimensional numerical
simulations of a magnetized cylindrical Taylor-Couette flow inspired
by laboratory experiment aiming to observe the magnetorotational
instability. 

The influence of the boundaries on the MRI has been
studied. We have shown that that the finite axial extent of the flow
deeply modifies the nature of the bifurcation. In the presence of
realistic boundaries, the MRI appears from a strongly imperfect
supercritical pitchfork bifurcation, leading to several predictions
for the observation of the instability in the laboratory. First, the
transition to MRI is expected to gradually emerge from the residual
recirculation driven by the top and bottom boundaries. Moreover, this
Ekman flow directly modifies the structure of the instability by
constraining the geometry of the mode, and different MRI modes can be
observed depending on the hydrodynamical background state. As a
consequence, the endcap rings of the Princeton experiment are not only
useful to reduce the Ekman recirculation, but they can also be used to
select different MRI modes.

In the second part of this study, we have reported the generation of
non-axisymmetric modes when the applied magnetic field is sufficiently
strong compared to the rotation (Elsasser number $>1$). Although they present some
similarities with the MRI, these modes are of a very different nature:
when the axial magnetic field is sufficiently strong, it generates a
new free shear layer in the flow, by extending the jump of angular
velocity between the inner and outer endcap rings into the bulk of the
flow. When the shear is sufficiently strong, this layer is
destabilized by non-axisymmetric modes of a Kelvin-Helmoltz type. It
is interesting to note that this new instability shares several
properties with the MRI: It transports angular momentum outward, it is
stabilized at strong applied field, and a critical magnetic field is
needed. However, these modes are inductionless, and therefore extend
to very small
$Rm$. This is an important difference from the classical MRI. 

To finish, we have compared our numerical simulations to experimental
results from the Princeton experiment.   Good agreement is
obtained at small $Rm$, where non-axisymmetric oscillations are also
observed in the experiment. The marginal stability curve, the geometry
of the mode, and the time evolution of these non-axisymmetric
oscillations are similar to our numerical simulations. This indicates
that the non-axisymmetric oscillations observed in the Princeton
experiment could be related to the destabilization of the free shear
layer (Shercliff layer) reported in the present numerical work.

To summarize, the present numerical study shows that the generation,
structure, and saturation of the MRI are strongly modified by the
presence of no-slip boundaries. In addition, new instabilities,
similar to the MRI but inductionless, are generated in the presence of
a strong magnetic field. This suggests that the observation of the
magnetorotational instability in a laboratory experiment could be
radically different from what is expected from local theory, or even
axially-periodic Taylor-Couette simulations. However, careful
comparisons between numerical simulations and experimental results
could lead to a clear identification of the magnetorotational
instability in the laboratory, and to a better understanding of the
angular momentum transport.

\begin{acknowledgments}
This work was supported by the NSF under grant AST-0607472, by the
NASA under grant numbers ATP06-35 and APRA08-0066, by the DOE under
Contract No. DE-AC02-09CH11466, and by the NSF Center for Magnetic
Self-Organization under grant PHY-0821899.  We have benefited from
useful discussions with E. Edlund, E. Spence and A. Roach.
\end{acknowledgments}


\end{document}